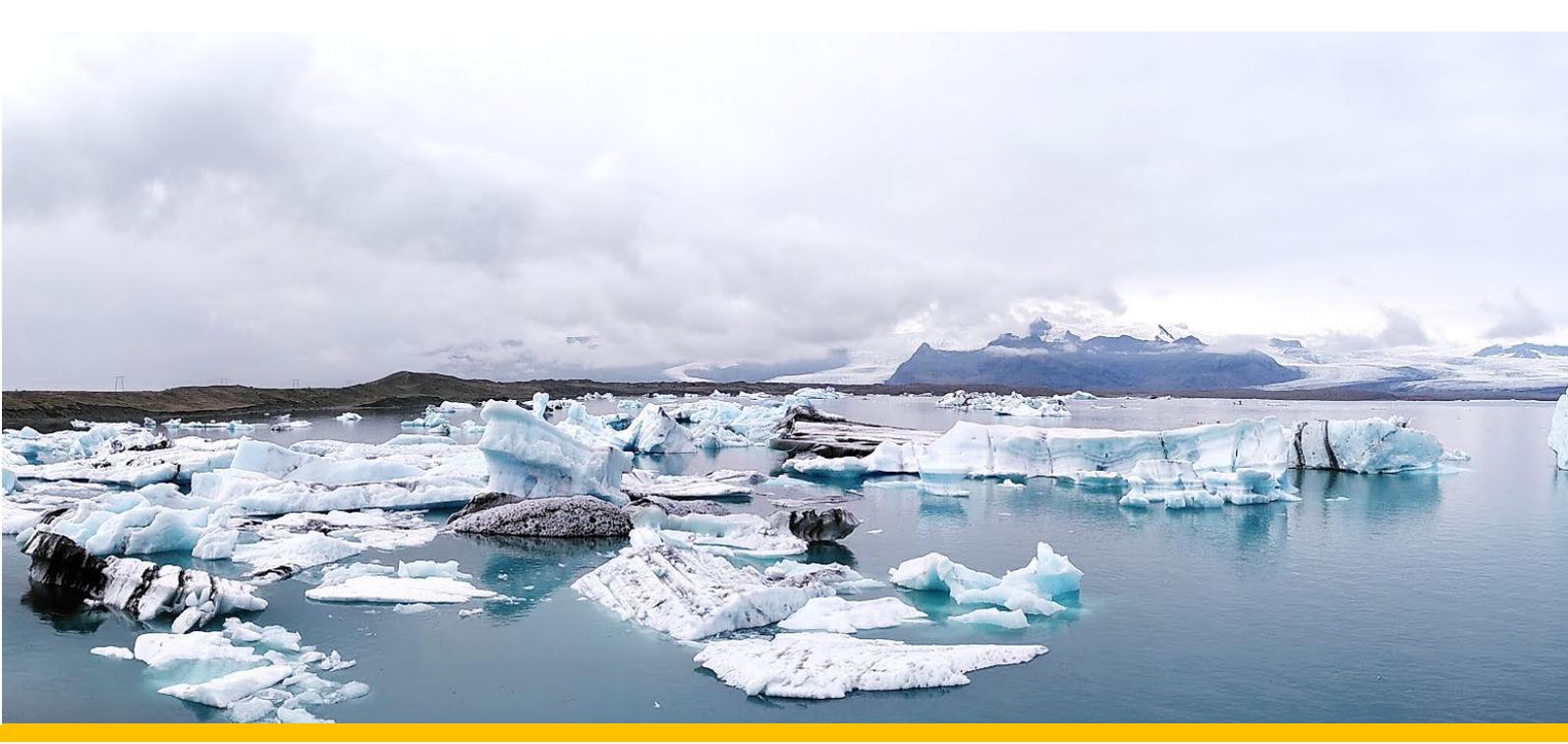

# Technologische Perspektive der digitalen Souveränität

**Blick auf die Schweiz, internationale Trends sowie Empfehlungen für die «Strategie Digitale Souveränität der Schweiz»**

Bericht zuhanden des Eidgenössischen Departements für auswärtige Angelegenheiten (EDA)

**Prof. Dr. Matthias Stürmer, Leiter Institut Public Sector Transformation**
BFH Wirtschaft, Brückenstrasse 73, 3005 Bern
matthias.stuermer@bfh.ch / +41 76 368 81 65

Bern, 12. Juni 2024

Departement Wirtschaft
Institut Public Sector Transformation

# Management Summary

**Der vorliegende Bericht zuhanden des Eidgenössischen Departements für auswärtige Angelegenheiten (EDA) leistet einen wissenschaftlichen Beitrag im Rahmen des Postulats 22.4411 «Strategie Digitale Souveränität der Schweiz» von Ständerätin Heidi Z'graggen. Der Bericht zeigt auf, was digitale Souveränität aus technologischer Perspektive bedeutet und welche Aktivitäten diesbezüglich in der Schweiz und im Ausland zurzeit realisiert werden. Ausserdem werden strategische Stossrichtungen und konkrete Empfehlungen für ein künftige «Strategie Digitale Souveränität der Schweiz» abgegeben.**

Im ersten Kapitel erläutert der Bericht, was unter «digitaler Souveränität» verstanden wird und wie sich der Begriff von ähnlichen Konzepten abgrenzt. Dazu wird eine im deutschsprachigen Raum verbreitete Definition analysiert, die sich sowohl auf den Umgang mit Daten als auch auf technologische Komponenten und IT-Systeme bezieht. Darüber hinaus wird aufgezeigt, wie sich die Ziele der digitalen Souveränität in die jahrzehntelange Entwicklung der Informatik einordnen lassen. Es wird der Bezug zu Hersteller-Abhängigkeiten, Open Source Software, Public Cloud Services und zu Cybersecurity hergestellt, um so die Zusammenhänge zwischen digitaler Souveränität und etablierten IT-Konzepten aufzuzeigen.

Das zweite Kapitel umfasst einen aktuellen Überblick, in welcher Art hierzulande digitale Souveränität bereits thematisiert wurde. Aus technologischer Sicht wichtig ist die Diskussion um die Swiss Cloud, die künftig als Swiss Government Cloud realisiert werden soll. Weiter spielen westschweizer Kantone eine zentrale Rolle in ihrem politischen Anliegen, die digitale Souveränität zu stärken, wozu sie 2022 und 2023 drei Grundlagenberichte verfassen liessen. Bereits im Jahr 2021 gab das EDA eine Studie zu Datenkolonialismus, Datensouveränität und digitaler Nachhaltigkeit in Auftrag und das Bundesamt für Kommunikation (BAKOM) ein Bericht zum «digitalen Service Public». Beide Publikationen betreffen Aspekte, die im Hinblick auf digitale Souveränität weiterhin aktuell sind. Auch bereits vorhanden ist das 2024 in Kraft getretene «Bundesgesetz über den Einsatz elektronischer Mittel zur Erfüllung von Behördenaufgaben» (EMBAG), das in verschiedener Hinsicht Grundlagen für die Umsetzung von digitaler Souveränität schafft.

Im dritten Kapitel wird aus globaler Perspektive beschrieben, wie in Deutschland, Frankreich, Indien, Lateinamerika, den USA, China, in der Europäischen Kommission, beim IKRK und bei den Vereinten Nationen die digitale Souveränität vorangetrieben wird. Neben konzeptionellen Aspekten werden insbesondere eine Vielzahl konkreter Vorhaben und Regulierungen betrachtet, die aufzeigen, dass im Ausland an vielen Stellen digitale Souveränität zurzeit wesentlich forcierter umgesetzt wird als in der Schweiz. So investieren insbesondere Deutschland und die Europäische Kommission aktuell Milliarden von Euro in technologische Initiativen und fördern so die digitale Souveränität in Europa.

Das vierte Kapitel zeigt strategische Stossrichtungen auf, die für die «Strategie Digitale Souveränität der Schweiz» grundlegende Impulse geben sollen. Einerseits sind dabei die technologischen Themenfelder Software, Daten, IT-Infrastruktur und künstliche Intelligenz relevant, die mit ihren unterschiedlichen Eigenschaften die digitale Realität abbilden. Andererseits zeigen sektorielle Betrachtungen auf das Finanzwesen, das Gesundheitswesen, den Verkehrssektor und den Bildungssektor auf, welche spezifischen Aspekte zu berücksichtigen sind und welche Gesetzgebungen und Massnahmen im In- und Ausland einen Einfluss auf die digitale Souveränität ausüben.

Im fünften Kapitel werden 13 ausführlich begründete Empfehlungen abgegeben, wie der Bund die digitale Souveränität in der Schweiz wirkungsvoll fördern könnte. Bei den Massnahmen wurde nach Möglichkeit eine Schätzung der Kostenfolgen und mögliche zuständige Behördenstellen angegeben.

Abschliessend muss festgestellt werden, dass die digitale Souveränität heute in der Schweiz stark eingeschränkt ist. Es gibt hierzulande einzelne Ansatzpunkte, wie die Kontrolle über die Daten hergestellt und die IT-Systeme eigenständig weiterentwickelt werden sollen, aber noch ist kein systematisches Vorgehen und keine konkrete Umsetzung sichtbar. Anders ist dies bei Nachbarstaaten wie Deutschland und Frankreich zu beobachten, bei denen digitale Souveränität regelmässig auf höchster Regierungsstufe diskutiert wird und die substanziell in die Unabhängigkeit ihrer Digitaltechnologien und IT-Infrastrukturen investieren. Auch die Schweiz hat zahlreiche Möglichkeiten, wirkungsvoll und kostengünstig die digitale Souveränität zu erhöhen, wenn sie sich an den vorliegenden Empfehlungen orientiert und Aktivitäten in diese Richtung realisiert.

# Inhaltsverzeichnis

# Danksagung

# Glossar

**Unterschiedliche Begrifflichkeiten bezüglich Souveränität in der digitalen Welt**

**Cyber-Souveränität** («cyber sovereignty»), **Internet-Souveränität** («internet sovereignty») oder **Netzwerk-Souveränität** («network sovereignty»): Fokus auf technische Kontrolle des im Internets (Netzneutralität, Blockieren von bestimmten Adressen, Steuerung des «Domain Name System» DNS und von Top Level Domains wie «.ch», Routing von Anfragen etc.) und Schutz der IT-Infrastruktur eines Landes, auch in Bezug auf militärische Sicherheit und territoriale Souveränität (Jensen, 2015; Zinovieva and Shitkov, 2023)

**Daten-Souveränität** («data sovereignty»): Fokus auf Souveränität bezüglich Daten-Management, Data Governance, Datenschutzrechte und internationaler Datenfluss (Pons, 2023)

**Digitale Souveränität** («digital sovereignty») oder **Technologische Souveränität** («technological sovereignty») oder **digitale Autonomie** («digital autonomy»): Fokus auf Souveränität in der Digitalisierung; umfasst Souveränität bezüglich Daten-Management, Souveränität im Internet (Cyberspace) und Souveränität der Informationen (Bundesministerium für Wirtschaft und Energie, 2018; Lambach and Oppermann, 2023); Kontrolle und eigenständige Entscheidungen bezüglich der IT-Infrastrukturen, Daten, Software und KI-Modelle

**Indigenous Data Sovereignty:** Souveränität bezüglich Daten indigener Menschen, Selbstbestimmung bezüglich Freigabe und Nutzung ihrer Daten (Taylor and Kukutai, 2016; Kukutai, 2023)

**Informations-Souveränität** («information sovereignty»): Souveränität des Staats bezüglich Umgang mit Informationen und Gewährleistung der Privatsphäre der Bevölkerung, oftmals eine juristische Perspektive (Polcak and Svantesson, 2017)

**Weitere Fachbegriffe in diesem Bericht**

**Datenkolonialismus** («data colonialism») oder **Digitalkolonialisierung** («digital colonialism»): Daten von Menschen und Staaten werden als digitale Ressourcen durch grosse IT-Unternehmen gesammelt und kontrolliert; Vergleich mit dem neuzeitlichen Kolonialismus, währenddem Menschen und Ressourcen im globalen Süden durch Länder des Nordens ausgebeutet wurden (Couldry and Mejias, 2019b, 2019a)

**Digitaltechnologien:** Software, Daten (inkl. Standards), künstliche Intelligenz, IT-Infrastruktur (inkl. Hardware)

**IT:** Informationstechnologien

**ICT:** Informations- und Kommunikationstechnologien

**Public Cloud Services:** öffentlich verfügbare Cloud Services, also Dienstleistungen von IT-Anbietern mit Servern in Rechenzentren

# 1 Einleitung

Das Thema «digitale Souveränität» hat in den vergangenen Jahren an Relevanz gewonnen, obwohl der Begriff schon weit über ein Jahrzehnt existiert (Bellanger, 2011; Gueham, 2017). Inhaltlich kann die Thematik in unterschiedlicher Breite und Tiefe betrachtet werden. Auch zeigen die diversen Begriffsvariationen wie Cyber-Souveränität, Internet-Souveränität, Netzwerk-Souveränität, Daten-Souveränität und Informations-Souveränität (siehe Glossar) auf, dass «digitale Souveränität» eine hohe Komplexität und Vielschichtigkeit beinhaltet – bereits das Begriffsverständnis und die Begriffsabgrenzungen stellen eine Herausforderung dar.

Im Bewusstsein der Historie, der Multidisziplinarität und der begrifflichen Mehrdeutigkeit fokussiert dieser Bericht auf die technologischen Aspekte der digitalen Souveränität: Software, Daten, IT-Infrastruktur und künstliche Intelligenz. Eine Auslegeordnung im Inland, internationale Beispiele sowie Sektor-spezifische Stossrichtungen und konkrete Empfehlungen zeigen auf, wie die digitale Souveränität in der Schweiz durch realistische und dennoch wirkungsvolle Massnahmen erhöht werden kann.

## 1.1 Definition digitale Souveränität

Der Begriff «digitale Souveränität» wird oft verwendet, aber unterschiedlich verstanden und definiert. Forschende haben 2023 im Rahmen einer systematischen Begriffsanalyse von 63 Publikationen zu digitaler Souveränität allein in Deutschland insgesamt sieben unterschiedliche Narrative identifiziert, welche die vielseitige Interpretation des Konzepts verdeutlicht: wirtschaftlicher Wohlstand, Sicherheit, die "europäische Lebensart", der moderne Staat, Datenschutz, Verbraucherschutz und demokratische Ermächtigung (Lambach and Oppermann, 2023).

In diesem Bericht soll «digitale Souveränität» gemäss der am deutschen Digital Gipfel 2018 vorgeschlagenen Definition verwendet werden (Bundesministerium für Wirtschaft und Energie, 2018). Diese Definition wurde unterdessen auch in zahlreichen weiteren Publikationen zu digitaler Souveränität angewendet (Pohle, 2020; Seifried and Bertschek, 2021; Itrich, Garloff and Kronlage-Dammers, 2022; Urban and Garloff, 2022), sodass in diesen Kreisen nun ein Konsens über die Begrifflichkeit besteht:

> *«Digitale Souveränität eines Staates oder einer Organisation umfasst zwingend die vollständige Kontrolle über gespeicherte und verarbeitete Daten sowie die unabhängige Entscheidung darüber, wer darauf zugreifen darf. Sie umfasst weiterhin die Fähigkeit, technologische Komponenten und Systeme eigenständig zu entwickeln, zu verändern, zu kontrollieren und durch andere Komponenten zu ergänzen.»*

Drei relevante Aspekte dieser Definition von digitaler Souveränität sind dabei hervorzuheben und werden deshalb nachfolgend erläutert.

### A) Abgrenzung zu «digitaler Selbstbestimmung» und «Datensouveränität»

Wichtig bei dieser Begriffsdefinition ist einerseits der Fokus auf den Handlungsspielraum von «Staaten und Organisationen». Es geht somit nicht um die Entscheidungsmöglichkeiten von Einzelpersonen, was meist als «digitale Selbstbestimmung» (Mollen and Haas, 2021) oder «informationelle Selbstbestimmung» (Goldacker, 2017) verstanden wird. Andererseits umfasst der Begriff «digitale Souveränität» in der obenstehenden Definition sowohl Daten als auch technologische Komponenten und Systeme. Damit grenzt sich dieses Begriffsverständnis auch gegenüber dem Konzept der «Datensouveränität» ab, das sich ausschliesslich auf Daten fokussiert (Laux and Wüst, 2022).

### B) Neben Daten auch digitale Infrastruktur und Technologie-Knowhow nötig

Auch bei dieser Definition der digitalen Souveränität steht die Datenhoheit und der Schutz vor unbefugtem Datenzugriff im Mittelpunkt, denn die «vollständige Kontrolle über die gespeicherten und verarbeiteten Daten» sowie die «unabhängige Entscheidung» über die Zugriffsrechte müssen vorhanden sein. Daten haben unbestritten die Bedeutung von «Öl für das 21. Jahrhundert» – ein universelle, wertvolle und strategische Ressource für Wirtschaft und Politik (Gueham, 2017).

Zur Erreichung von digitaler Souveränität werden aber noch weitere Elemente benötigt (siehe auch Thumfart, 2022, Seite 4): Als zweiter, wichtiger Aspekt für digitale Souveränität sind die «technologischen Komponenten und Systeme» erwähnt. Dies bezieht sich auf die benötigte digitale Infrastruktur wie Software und Hardware, um die Datenverarbeitung tatsächlich «eigenständig» realisieren zu können. Und drittens ist neben den virtuellen Daten und Programmen und der physischen IT-Infrastruktur insbesondere auch die «Fähigkeit» notwendig, um diese komplexen Systeme zu kontrollieren und weiterzuentwickeln. Damit ist klar, dass für die Umsetzung der digitalen Souveränität insbesondere auch fähige Menschen mit dem entsprechenden Technologie-Knowhow benötigt werden.

*C) Geografische Datenspeicherung nicht erwähnt, Wahlmöglichkeit im Zentrum*

Interessant bei diesem Verständnis von digitaler Souveränität ist ausserdem die Tatsache, dass die geografische Kontrolle nicht erwähnt ist. Es spielt somit keine Rolle, ob die physischen Server und Daten im selben Land liegen oder nicht, sondern es geht vielmehr um die eigenständigen, unabhängigen Entscheidungsmöglichkeiten. Gemäss dieser Begriffsdefinition bedingt die Möglichkeit zur eigenständigen Kontrolle und Veränderung von «technologischen Komponenten und Systemen» somit auch die Wahl zwischen verschiedenen alternativen, gleichwertigen Technologien und Anbietern.

## 1.2 Ausgangslage und grundlegende Feststellungen zu digitaler Souveränität

Die digitale Transformation wird im Wesentlichen durch neue technologische Möglichkeiten vorangetrieben, die von IT-Unternehmen entwickelt werden. In den letzten zehn Jahren haben innovative Produkte wie das Smartphone oder praktische Lösungen wie Cloud Computing die Kommunikation und die Zusammenarbeit in der Wirtschaft und im öffentlichen Sektor erleichtert. Gleichzeitig ist es typischerweise Teil des Geschäftsmodells dieser IT-Anbieter, eine möglichst hohe Kundenbindung zu schaffen, in dem sie ihre Anwendenden durch den sogenannten «Vendor Lock-In» abhängig machen (Opara-Martins, Sahandi and Tian, 2016).

In der Folge haben die IT-nutzenden Behörden und anderen Organisationen kaum mehr die Möglichkeit für einen Anbieterwechsel, da sich sowohl ihre Anwendungen als auch ihre darin enthaltenen Daten unter Kontrolle der IT-Unternehmen befinden. Die Konsequenzen sind «eingeschränkte Informationssicherheit, rechtliche Unsicherheit, unkontrollierbare Kosten, eingeschränkte Flexibilität und fremdgesteuerte Innovation» (BfIT, IT-Planungsrat, and IT-Rat, 2020). Und wenn sich schliesslich die Industrien ganzer Nationen in die Abhängigkeit von wenigen ausländischen IT-Anbieter begeben, kann dies zu Verlust von internationaler Wettbewerbsfähigkeit eines Landes führen (Kar and Thapa, 2020).

Eine Antwort auf diese wachsende Abhängigkeit von IT-Anbietern stellt die digitale Souveränität dar. Dieses zeigt auf, wie durch die Entwicklung von technologischen Alternativen und durch den Aufbau von Digitalkompetenzen der Mitarbeitenden realistische Wechselmöglichkeiten geschaffen werden können (Kagermann, Streibich, and Suder, 2021).

## 1.3 System- versus Hersteller-Abhängigkeiten

Die Thematik von Abhängigkeiten in der IT ist altbekannt. Schon in den 90er-Jahren hat Shane Greenstein auf die Herstellerabhängigkeit von IBM bei den Mainframe-Computer hingewiesen (Greenstein, 1997). Dieser «Vendor Lock-In» wurde durch hohe Wechselkosten («Switching Cost») erzeugt, die notwendig wären, um auf ein ähnliches Produkt umzusteigen. So hatten die Hersteller seit jeher einen Anreiz inkompatible, proprietäre System zu entwickeln um die Wechselkosten der Anwendenden in die Höhe zu treiben (Katz and Shapiro, 1985). Die Nutzenden von solchen IT-Lösungen begaben sich wiederum durch frühere Entscheidungen in sogenannte «Pfadabhängigkeiten», aus denen sie kaum mehr entfliehen konnten (Liebowitz and Margolis, 1995).

In der Praxis wird nun oftmals von einem «Lock-In» gesprochen, ohne zwischen zwei wesentlich unterschiedlichen Arten von Abhängigkeiten zu differenzieren: Der Abhängigkeit von der IT-Lösung (IT-System, Software bzw. digitales Produkt) versus der Anbieter- bzw. Herstellerabhängigkeit. Diese verschiedenen Eigenschaften der Abhängigkeiten werden nachfolgend erläutert.

*System-Abhängigkeiten*

Es liegt in der Natur der Sache und ist auch Ziel von IT-Projekten, die eingesetzten Technologien möglichst eng mit den vorhandenen IT-Systemen über Schnittstellen oder kompatible Dateiformate zu integrieren. So werden zwar technische Abhängigkeiten geschaffen, die jedoch dem Zweck von interoperablen Lösungen entsprechen und damit eine einheitliche Gesamtlösung erzielen. Des Weiteren ist es notwendig, dass die IT-Anwendenden die eingesetzten Programme gut kennenlernen, sich an die Funktionalitäten und technischen Möglichkeiten gewöhnen und so entsprechendes Nutzungs-Knowhow aufbauen um letztlich eine hohe Technologieakzeptanz zu erreichen (Davis, 1989; Venkatesh et al., 2003).

Durch die Anpassung von Abläufen und organisationale Strukturen wird die IT-Lösung eng in die Organisation eingebunden, sodass organisationale Abhängigkeiten von den eingesetzten IT-Systemen entstehen.

*Hersteller-Abhängigkeiten*

Wesentlich andere Arten von Abhängigkeiten werden durch die Hersteller und Dienstleistungsanbieter der IT-Lösungen geschaffen. Diese Unternehmen generieren einerseits rechtliche Abhängigkeiten durch die Gewährung von Nutzungslizenzen («End User License Agreements» EULA) und weiteren Vertragsbedingungen, welche die Firmen aufgrund ihres Urheberrechts an den Software-Produkten vorgeben können (Zhu and Zhou, 2012). Andererseits schaffen die Hersteller und Anbieter Knowhow-Abhängigkeiten, da ihre Mitarbeitenden das Wissen und die Erfahrung für den Betrieb, die Fehlerbehandlung und die Weiterentwicklung der entsprechenden IT-Lösungen besitzen.

Und letztlich werden auch nicht zu unterschätzende psychologische Abhängigkeiten mittels Marken («Brands») von Firmen und Produkten sowie über Bekanntheit und Verbreitung von entsprechenden IT-Lösungen aufgebaut. So existiert in der Informatik das bekannte Sprichwort «Nobody ever got fired for buying IBM» (Silic and Back, 2016), wonach sich keine entscheidungstragende Person Sorgen machen muss, wenn sie etablierte Produkte von bekannten Herstellern kauft. Werden hingegen weniger verbreitete Lösungen wie bspw. Open Source Software eingesetzt, wird dies als unnötiges Risiko angesehen, das bei einem Scheitern des Projekts ins Rampenlicht gerückt wird und so zu beruflichen Nachteilen führen kann.

*Stabile digitale öffentliche Infrastruktur*

In Bezug auf digitale Souveränität bedeutet dies, dass die vollständige Wahlfreiheit bei den möglichen IT-Lösungen eingeschränkt werden muss, um den erfolgreichen Einsatz von Digitaltechnologien zu erreichen. Denn nur wenn sich Mitarbeitende und Organisationen auf die fachlichen Eigenschaften der gewählten Produkte einlassen und diese über technische Schnittstellen ideal integrieren, können diese IT-Lösungen effizient und zielgerichtet im Arbeitsalltag eingesetzt werden. Somit ist es nicht zielführend, die Technologien häufig zu wechseln, da so deren wirtschaftlicher Nutzen verloren geht. Es braucht eine möglichst stabile, modulare digitale öffentliche Infrastruktur, die laufend weiterentwickelt werden kann. Im öffentlichen Sektor wird diesbezüglich auch von der «Digital Public Infrastructure» gesprochen (Zuckerman, 2020; Deloitte, 2023; ITU and UNDP, 2023).

*Verbesserte digitale Souveränität beim Einsatz von Open Source Software*

Währenddem die digitale Infrastruktur stabil bleibt, kann digitale Souveränität dennoch durch die Reduktion von Hersteller-Abhängigkeiten erreicht werden. Voraussetzung dazu ist die Wechselmöglichkeit bei IT-Firmen, die Dienstleistungen für gut integrierte IT-Lösungen anbieten. Erreicht wird dies durch die Reduzierung von rechtlichen Abhängigkeiten (proprietärer Lizenzen) und den Aufbau von Technologie-Knowhow bei den Anwendenden.

Zentral dabei ist der Einsatz von Open Source Software bei der digitalen öffentlichen Infrastruktur. Solange mehrere IT-Firmen die bestehenden IT-Systeme warten und weiterentwickeln können, besteht ein gesunder Wettbewerb und die Hersteller-Abhängigkeit wird minimiert. Dies erhöht die Stabilität der digitalen Infrastruktur, die von unterschiedlichen Dienstleistungsanbietern gewartet werden kann.

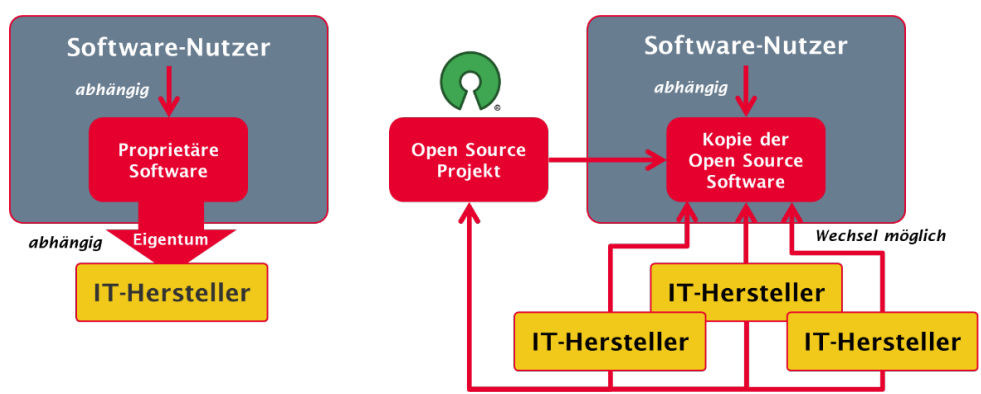

Abbildung 1: Unterschiedliche Abhängigkeiten bei proprietärer und bei Open Source Software (eigene Darstellung)

Wie in zahlreichen Berichten zu «Digital Public Infrastructure» erläutert, reduziert der Einsatz von Open Source Software die Hersteller-Abhängigkeit und erhöht damit die digitale Souveränität des öffentlichen Sektors (O'Neil and Rasul, 2021; Digital Public Goods Alliance, 2022; Meier and Arensen, 2023). Dies war auch bereits 2019 eine der Schlussforderungen eines ausführlichen Berichts der Weltbank über das Potenzial von Open Source Software in der Behörden-Informatik (World Bank, 2019). Dementsprechend ist eine der Kernempfehlungen dieses Berichts der systematische Einsatz von Open Source Software in der Informatik des öffentlichen Sektors (siehe «Massnahme 4: Förderung von Open Source Software im öffentlichen Sektor»).

### 1.4 Trade-off zwischen Public Cloud Services und digitaler Souveränität

Eine weitere technologische Betrachtungsweise ergibt sich aus der heute relevanten Diskussion über die Vor- und Nachteile bei der Nutzung von «Public Cloud Services». Darunter werden einerseits Angebote von grossen, internationalen Technologiekonzernen wie Amazon, Microsoft oder Google verstanden, die als sogenannte «Hyperscalers» eine Vielzahl an verschiedenen Cloud Services anbieten. Andererseits betrifft es auch Unternehmen wie SAP, die ihre Lösungen künftig über eine eigene «Enterprise Cloud» an ihre Kunden vertreiben möchten (Bögelsack et al., 2022). Die folgende Grafik illustriert den klassischen Trade-Off zwischen Public Cloud Services und digitaler Souveränität:

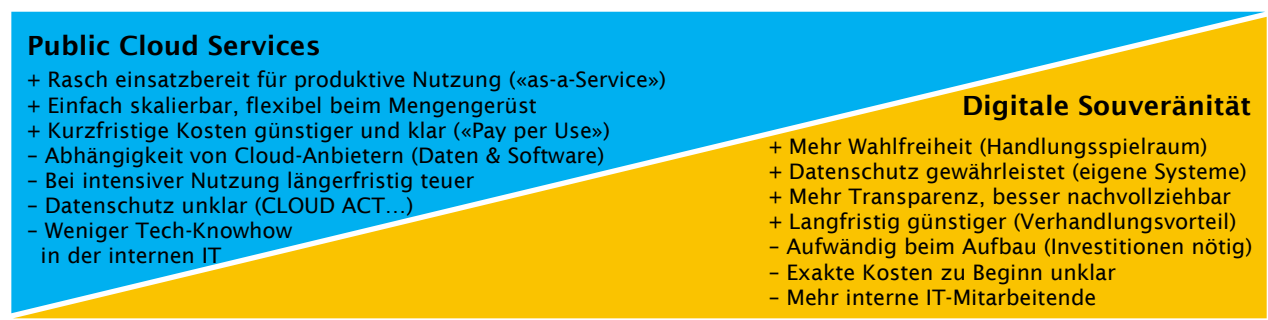

Abbildung 2: Trade-Off zwischen Public Cloud Services und digitaler Souveränität (eigene Darstellung)

*Vor- und Nachteile von Public Cloud Services*

Obwohl technisch gesehen auch bei der «Public Cloud» im Hintergrund immer noch Rechenzentren mit vielen Servern stehen, die sich in der Kontrolle der jeweiligen Anbieter befinden, hat sich in der Praxiswelt die Begrifflichkeit des «Cloud Computing» heute breit etabliert. Das Wertversprechen der Cloud Services umfasst dabei die technische Einfachheit, die Skalierbarkeit und die (zunächst) niedrigen Kosten. Gleichzeitig wird eine Abhängigkeit der Cloud-Anbieter in Kauf genommen, welche sowohl die für die Cloud-Dienste notwendige Software als auch die IT-Infrastruktur besitzen und ausserdem weitgehend die Datenhaltung kontrollieren. Bezüglich Preise können je nach Anwendungsfall die Cloud Services bei intensiver Nutzung längerfristig jedoch teurer sein als selbst betriebene Systeme (Vaughan, 2023), der Datenschutz ist aufgrund des CLOUD ACTs unklar (Abed and Chavan, 2019) und in der internen Informatik wird weniger Knowhow aufgebaut.

*Vor- und Nachteile von digital souveränen IT-Systemen*

Das Konzept der «digitalen Souveränität» bietet in vielerlei Hinsicht das Gegenteil dazu. Werden IT-Systeme auf eigenen Servern («On-Premises») betrieben, besteht mehr Wahlfreiheit und dadurch auch ein grösserer Handlungsspielraum bei den IT-Anbietern für Support und Wartung. Auch ist der Datenschutz gewährleistet, da die IT-Systeme der Kontrolle der Datenbesitzer unterliegen. Die Technologien wie bspw. die Datenspeicherung sind transparenter nachvollziehbar. Und dank der höheren Wahlfreiheit entsteht ein Verhandlungsvorteil gegenüber den Anbietern wodurch niedrigere Kosten entstehen. Allerdings sind Investitionen in die Hardware und das Personal notwendig um eine eigene IT-Infrastruktur aufzubauen, die genauen Kosten sind zu Beginn schwierig abzuschätzen und es braucht mehr internes IT-Knowhow.

*Trade-Off kann mit «digital souveränen Clouds» gelöst werden*

Auch wenn die oben dargestellte Abbildung den Anschein macht, dass es stets einen Kompromiss zwischen Public Cloud Services und digitaler Souveränität braucht, wurden in den letzten Jahren Ansätze entwickelt, bei denen die positiven Aspekte von beiden Seiten genutzt werden können: Die Stärken von

Public Cloud Anbietern mit den Vorteilen digitaler Souveränität lassen sich kombinieren, indem die IT-Infrastruktur mittels Open Source Cloud Technologien betrieben wird. Dabei kann dieser Software-Stack auf eigenen Servern («On-Premises») oder auch auf standardisierten «Infrastructure-as-a-Services» (IaaS) betrieben werden.

Den Ansatz der selber kontrollierten Cloud-Infrastruktur verfolgt im Rahmen von Gaia-X der «Sovereign Cloud Stack», der auf einer standardisierten Zusammenstellung von zahlreichen Open Source Lösungen basiert (Urban and Garloff, 2022). Das Vorhaben der Open Source Business Alliance (OSBA, 2023) wird von der deutschen Regierung politisch und finanziell mit über 100 Millionen Euro gefördert (Caspers, 2022; Puppe, 2023). Mit diesem «Sovereign Cloud Stack» sollen die Anwendenden im Bedarfsfall relativ einfach den Anbieter wechseln können, denn die Herstellerabhängigkeit lässt sich durch die Verwendung von interoperablen Cloud Services vermeiden. Verwenden die in der Public Cloud betriebenen Applikationen standardisierte Schnittstellen, können die Anbieter der IT-Infrastruktur bei Bedarf mit wenig Aufwand gewechselt werden (Michels, Millard and Walden, 2023). Damit besteht nur noch eine Abhängigkeit zu den eingesetzten Open Source Produkten, jedoch nicht mehr zu den jeweiligen Herstellern und Betreibern der Public Cloud Infrastruktur. Diese können und sollen sich weiterhin im Wettbewerb um leistungsfähige Rechenzentren und Server konkurrieren, der bisherige «Vendor Lock-In» fällt jedoch weg.

*Aus der Public Cloud zurück zu «On-Premises»*

Es gibt bereits heute Akteure, die aufzeigen, dass es technisch relativ einfach machbar ist und wirtschaftlich sinnvoll sein kann, aus der Public Cloud von «Hyperscalers» hin zu eigener Server-Hardware und einem eigenen Open Source Cloud Stack zu wechseln (Baset, 2012; Barkat, dos Santos and Nguyen Ho, 2014; Kumar *et al.*, 2014). So hat die Firma 37signals im Jahr 2023 ihre gesamten Kundenapplikationen aus der Public Cloud auf selber betriebene Server migriert (Heinemeier Hansson, 2023a). Zuvor bezahlte die Firma jährlich USD 3.2 Millionen an Cloud-Gebühren. Durch eine einmalige Investition von USD 600'000 in Server-Hardware will die Firma in den nächsten fünf Jahren rund USD 7 Millionen einsparen (Heinemeier Hansson, 2023b).

Interessanterweise benötigt 37signals dazu nicht mehr oder andere Personalressourcen, denn auch für den Applikationsbetrieb in der Public Cloud braucht es gut ausgebildete Fachleute. Der grösste Nutzen der Public Cloud Services war die einfache und rasche Aufschaltung von zusätzlichen virtuellen Servern und anderer Cloud-Ressourcen, was nur wenige Sekunden benötigt. Der Einkauf und die Installation von eigenen, physischen Servern dauert mehrere Wochen, dafür entstehen danach deutlich günstigere Betriebskosten. Die Sicherheit leidet nicht darunter, da im Public Cloud Betrieb auf dieselben Aspekte geachtet werden muss, wie wenn 37signals nun eigene Server hostet (Heinemeier Hansson, 2023a).

### 1.5 Bezug von digitaler Souveränität zum Thema Cybersecurity

Cybersecurity umfasst verschiedene Aspekte wie Sicherheit vor Datendiebstahl und Betriebsstabilität im Sinne von Ausfallsicherheit. Es versteht sich daher von selbst, dass kritische IT-Infrastrukturen und sensible Daten mit höchsten Anforderungen an die Cybersicherheit zu behandeln sind.

In der Praxis wird gelegentlich die Frage gestellt, ob Public Cloud Services von grossen IT-Unternehmen («Hyperscalers») oder digital souverän betriebene IT-Infrastruktur des öffentlichen Sektors sicherer sind. Eine abschliessende, auf wissenschaftlichen Fakten basierende Beurteilung erscheint schwierig. Es lassen sich jedoch objektiv Stärken und Schwächen der jeweiligen Varianten darstellen:

|  | **Public Cloud Services von Hyperscalern** | **Digital souveräne IT-Infrastruktur des öffentlichen Sektors** |
|---|---|---|
| **Stärken** | • Mehr Sicherheitspersonal, grössere Skaleneffekte<br>• Technische Verantwortung beim IT-Anbieter | • Klarheit über verwendete IT-Komponenten<br>• Keinen Einfluss ausländischer Staaten |
| **Schwächen** | • Grosses Angriffspotenzial, hohes Klumpenrisiko<br>• Wenig Transparenz bei Problemfällen | • Weniger finanzielle und personelle Ressourcen<br>• Technische Verantwortung bei der Verwaltung |

Tabelle 1: Vergleich von Stärken und Schwächen bezüglich Sicherheit von Public Cloud versus digitaler Souveränität

Es kann somit nicht generell festgestellt werden, ob Public Cloud Services von Hyperscalern oder die digital souveräne IT-Infrastruktur des öffentlichen Sektors mehr Sicherheit bieten. Grosse IT-Unternehmen verfügen zwar über mehr Sicherheitspersonal und können von grösseren Skaleneffekten profitieren. Gleichzeitig wächst aber auch das Angriffspotenzial und somit das Klumpenrisiko, wenn etwas schief geht.

So wurde bspw. im Sommer 2023 ein Master-Key von Microsoft gestohlen, der potenziell Zugang zu grossen Datenbeständen auf den Cloud-Lösungen von Microsoft gewährt hat (Schmidt, 2023; Tamari, 2023). Kritisch dabei war auch, dass Microsoft bis zum jetzigen Zeitpunkt nicht transparent über die Vorkommnisse informiert hat und entsprechend unklar ist, welcher Schaden tatsächlich entstanden ist. Und bereits im Januar 2024 wurde Microsoft erneut von Hackern angegriffen, dieses Mal von der russischen Gruppe «Midnight Blizzard», die sich offenbar Zugriff auf Emails und Quellcode verschaffen konnten (Microsoft, 2024; Stöckel, 2024). Aufgrund dieser Vorfälle prüfen nun US-amerikanische Behörden einen teilweisen Wechsel in die Cloud-Systeme von Google und Amazon um ihre Abhängigkeiten von Microsoft zu reduzieren (Bakolia, 2024; derStandard, 2024).

Auf der anderen Seite sind digitale souveräne IT-Infrastrukturen von öffentlichen Akteuren ebenfalls Cyberattacken ausgesetzt, müssen jedoch mit eigenen, beschränkten Mitteln geschützt werden. Dafür ist einfacher nachvollziehbar, welche technischen Komponenten verwendet werden. Zudem ist sichergestellt, dass keine ausländischen Staaten oder Firmen Zugriff auf die Daten und Systeme erhalten. Die Gefahr von Hintertüren in den Applikationen für ungewollten Datenabfluss (sogenannte «Backdoors») kann somit wesentlich reduziert werden.

# 2 Aktuelle Situation der digitalen Souveränität in der Schweiz

Bis zum heutigen Zeitpunkt hat die Diskussion in der Schweiz bezüglich digitaler Souveränität erst beschränkt stattgefunden. Massgebend ist das im Dezember 2022 eingereichte und im März 2023 durch den Bundesrat gutgeheissene Postulat 22.4411 «Strategie Digitale Souveränität der Schweiz» von Ständerätin Heidi Z'graggen. Darin wird der Bundesrat aufgefordert, eine Begriffsdefinition festzulegen, eine übergeordnete und umfassende Strategie zu präsentieren, den gesetzgeberischen Handlungsbedarf aufzuzeigen, Prioritäten und Massnahmen zu bestimmen, einen Zeitplan zu definieren und die notwendigen Mittel bereitzustellen, um dringende und erfolgsversprechende Projekte zur Stärkung der digitalen Souveränität rasch umzusetzen.

### 2.1 Swiss (Government) Cloud

Weitgehend unabhängig von diesem Vorstoss, aber inhaltlich ähnlich, hat die Schweizer Politik in den vergangenen Jahren mehrfach die Forderung nach einer «Swiss Cloud» geäussert. So kam zwar der Bundesrat 2020 zum Schluss, dass aktuell kein Bedarf an einer Schweizer Cloud-Lösung bestehe (ISB, 2020), aber dennoch u.a. die internationale Vernetzung mit der EU bezüglich Gaia-X und digitaler Souveränität gepflegt werden soll. Im September 2021 reichte Nationalrätin Isabelle Moret die parlamentarische Initiative 21.495 «Cybersicherheit. Schaffung einer eigenständigen digitalen Infrastruktur und Erarbeitung von Standards im Sicherheitsmanagement» ein, die den Aufbau von Schweizer Cloud-Diensten verlangte. Dem Vorstoss wurde in der nationalrätlichen sicherheitspolitischen Kommission zugestimmt, jedoch in der ständerätlichen Kommission abgelehnt.

Eine ähnlich lautende Motion 23.3002 «Mehr Sicherheit bei den wichtigsten digitalen Daten der Schweiz» wurde im Januar 2023 von Ständerat Josef Dittli als Kommissionsmotion eingereicht, durch den Bundesrat zur Annahme empfohlen und anschliessend im National- und Ständerat angenommen. Der Vorstoss greift somit das politische Anliegen der «Swiss Cloud» wieder auf, fokussiert sich dieses Mal jedoch auf die Bereitstellung von Cloud-Services ausschliesslich für den öffentlichen Sektor (inside-it.ch, 2023).

Gleichzeitig ist auch das Bundesamt für Informatik und Telekommunikation (BIT) daran, das Thema «Swiss Government Cloud» (SGC) neu voranzutreiben (Koller, 2023). In Koordination mit dem Bereich Digitale Transformation und IKT-Lenkung (DTI) der Bundeskanzlei soll das BIT in den nächsten Jahren Cloud-Dienste für die Bundesstellen, Kantone und weitere öffentliche Organisationen anbieten. Der Bundesrat hat dazu am 21. Mai 2024 die «Botschaft zu einem Verpflichtungskredit zum Aufbau einer Swiss Government Cloud» verabschiedet (Bundesrat, 2024). Das Vorhaben sieht vor, in den Jahren 2025 bis 2032 insgesamt CHF 319.4 Millionen in den Aufbau einer «Hybrid-Multi-Cloud-Infrastruktur» bestehend aus den Bereichen Public Cloud, Public Cloud On-Prem und Private Cloud On-Prem zu investieren.

## 2.2 Kantone in der Romandie wollen bezüglich digitaler Souveränität zusammenarbeiten

Im Mai 2023 haben die Kantone Freiburg, Genf, Jura, Neuenburg, Tessin, Waadt und Wallis bekannt gegeben, dass sie als «Conférence latine des directrices et directeurs cantonaux du numérique» (CLDN) zusammenarbeiten wollen um die digitale Souveränität voranzutreiben (CLDN, 2023). Als Grundlagenarbeit haben die Westschweizer Kantone drei Studien in Auftrag gegeben, welche digitale Souveränität aus technischer, rechtlich-ökonomischer und ethischer Perspektive untersucht haben (Savoy *et al.*, 2023; Benhamou, Bernard and Durand, 2023; Rochel, 2022).[1] Wichtig ist den Kantonen die Schaffung von geeigneten Rahmenbedingungen bezüglich Rechtsgrundlagen und Datenpolitik. Auch wollen sie sich der geplanten «Swiss Government Cloud» des Bundes anschliessen, um Synergien auf interkantonaler und eidgenössischer Ebene zu erschliessen.

## 2.3 Studie zu Datenkolonialismus, Datensouveränität und digitaler Nachhaltigkeit

Im Jahr 2021 beauftragte das Eidgenössische Aussendepartement die Universität Bern mit der Realisierung einer Studie zu Datenkolonialismus, Datensouveränität und digitaler Nachhaltigkeit (Stürmer, Nussbaumer and Stöckli, 2021). Darin erläuterten die Autorinnen und Autoren die Risiken, Konsequenzen und Empfehlungen um Abhängigkeiten von grossen IT-Unternehmen zu reduzieren. Eine Schlüsselerkenntnis dabei war, dass der öffentliche Sektor nicht nur mehr physische Digitaltechnologien aufbauen soll, sondern auch mehr Fachleute mit technischen Fähigkeiten für den Betrieb und die Weiterentwicklung von IT-Infrastrukturen, Software-Applikationen und Datenverarbeitung benötigt, wenn die Herstellerabhängigkeiten verkleinert werden sollen. Im Rahmen einer aufgezeichneten Online-Konferenz wurden die Erkenntnisse zusammen mit befragten Expertinnen und Experten im Mai 2021 präsentiert (Universität Bern, 2021).

## 2.4 Studie zur Rolle des Staates im digitalen Zeitalter im Kontext des «digitalen Service Public»

Im Rahmen des Postulats 19.3574 «Offensive für einen digitalen Service public» von Nationalrätin Min Li Marti verfasste die Berner Fachhochschule 2022 eine Studie (Gees et al., 2022) im Auftrag des Bundesamt für Kommunikation (BAKOM). Der Bericht behandelte auch die Frage, welche Leistungen der Staat im digitalen Zeitalter der Bevölkerung und der Wirtschaft erbringen soll. Dabei wird unter anderem thematisiert, wie der «digitale Service public» einen Beitrag zur Erhöhung der digitalen Souveränität schaffen kann. Beispielsweise wird argumentiert, dass das Angebot der geplanten staatlichen elektronischen Identität (E-ID) und der dazugehörigen Vertrauensinfrastruktur (Bundesrat, 2023c) die technologischen Fähigkeiten des Staates erhöht, die Entscheidungsmöglichkeiten der Bevölkerung stärkt und damit die digitale Souveränität der Schweiz verbessert.

## 2.5 Möglichkeiten des EMBAG zur Reduktion von Herstellerabhängigkeiten

In indirektem Zusammenhang mit digitaler Souveränität wurde im März 2023 das «Bundesgesetz über den Einsatz elektronischer Mittel zur Erfüllung von Behördenaufgaben» (EMBAG) verabschiedet (Bundesversammlung, 2023). Darin sieht Artikel 9 «Open Source Software» vor, dass der Bund künftig eigens entwickelte oder durch Externe realisierte Software «by default» unter einer Open Source Lizenz veröffentlicht (Stürmer, 2023). Auch besteht nun eine gesetzliche Grundlage, dass Bundesstellen ergänzende Dienstleistungen erbringen können, die zur besseren Integration, Wartung, Sicherheit und Support dienen. Mittels Aufbau von entsprechenden Open Source Communities kann der Bund so für seine Individualentwicklungen einen Markt von Dienstleistern aufbauen und dadurch die Herstellerabhängigkeiten reduzieren. Das EMBAG ist per 1. Januar 2024 in Kraft getreten.

# 3 Internationale Perspektive der digitalen Souveränität

Während das Thema «digitale Souveränität» in der Schweiz erst seit wenigen Jahren politisch und fachlich reflektiert wird, ist die Diskussion und die Umsetzung durch konkrete Massnahmen in anderen Ländern schon viel weiter fortgeschritten. Dabei wird stets die geopolitische Bedeutung der digitalen Souveränität hervorgehoben, wie sie unterdessen in einer Vielzahl von Ländern vorangetrieben wird (Glasze et al., 2023). So spielt digitale Souveränität insbesondere auch für China und Russland eine wichtige, sicherheitspolitische Rolle in der Abgrenzung zu den USA (Thumfart, 2022).

---

[1] Der Bericht über die rechtlich-ökonomischen Aspekte ist auch auf Englisch in einer Fachzeitschrift publiziert worden (Benhamou, Bernard and Durand, 2023a).

Die im folgenden Kapitel erläuterten Entwicklungen und Massnahmen aus anderen Staaten zeigen auf, wie die Wahrnehmung von digitaler Souveränität im Ausland ist und in welche Richtung es auch in der Schweiz weitergehen könnte. Neben ausgewählten europäischen Ländern werden insbesondere auch Aktivitäten bezüglich digitaler Souveränität in grossen Nationen wie den USA, Indien und China sowie in Lateinamerika und in multilateralen Organisationen wie der EU, dem IKRK und den Vereinten Nationen betrachtet. In vielen anderen Ländern der Welt (Australien, Vietnam, Südafrika, Russland etc.) wird das Thema digitale Souveränität ebenfalls behandelt, wie ein Bericht von der Internet Society illustriert (ISOC, 2022).

## 3.1 Deutschland

Deutschland gilt als eines der führenden Länder bezüglich Diskussion und Realisierung von «digitaler Souveränität». Zahlreiche Wissenschaftsorganisationen wie das Fraunhofer Institut (Goldacker, 2017; Edler et al., 2020; Kreutzer and Vogelsang, 2022), «acatech - Deutsche Akademie der Technikwissenschaften» (Kagermann and Wilhelm, 2020; Kagermann, Streibich, and Suder, 2021) oder die Konrad-Adenauer Stiftung (Pohle, 2020) haben umfassende Studien veröffentlicht. Eine vertiefte Zusammenarbeit mit Frankreich zu digitaler Souveränität, insbesondere zu «Open Source Large Language Models», wurde erst im Februar 2024 angekündigt (BMI, 2024). Konkret wurden in den letzten zwei Jahren zahlreiche operative Schritte unternommen, die technologische Realisierung von digitaler Souveränität in die Praxis umzusetzen. Nachfolgend sind fünf Initiativen kurz umschrieben.

### *Sovereign Cloud Stack*

Wie im Abschnitt 1.4 «Trade-off zwischen Public Cloud Services und digitaler Souveränität» bereits ausführlich beschrieben, bietet der «Sovereign Cloud Stack» eine standardisierte IT-Lösung basierend auf Open Source Cloud Technologien an. Damit können in bestehenden Rechenzentren «Infrastructure-as-a-Service» (IaaS) und «Container-as-a-Service» (CaaS) Systeme aufgebaut und betrieben werden, die als Grundlage für beliebige Cloud-Anwendungen dienen können (Urban and Garloff, 2022). Der deutsche Staat förderte die «Sovereign Cloud Stack» Initiative mit knapp 15 Millionen Euro (Witmer-Goßner, 2021).

### *Zentrum für digitale Souveränität (ZenDiS)*

Basierend auf strategischen Grundlagen verschiedener staatlicher Gremien (BfIT, IT-Planungsrat, and IT-Rat, 2020; IT-Planungsrat, 2021) haben Bund, Länder und Kommunen in Deutschland per Dezember 2022 das so genannte «Zentrum für digitale Souveränität» (ZenDiS) gegründet (Der Beauftragte der Bundesregierung für Informationstechnik, 2022). Dieses Zentrum hat zahlreiche Aufgaben zur Förderung von digitaler Souveränität und Open Source Software in Deutschland, unter anderem die Priorisierung von Vorhaben, die Entwicklung von Guidelines, die Beratung von öffentlichen Stellen, die Weiterentwicklung von Open Source Lösungen, Kooperationen mit Communities, Kommunikation und Aufklärung der öffentlichen Verwaltung zu Open Source Software etc. Im Jahr 2023 standen für den Aufbau rund 50 Millionen Euro zur Verfügung, 2024 sind aufgrund von Kürzungen im Bundeshaushalt noch rund 25 Millionen Euro vorgesehen (Kindermann, 2023).

### *Open CoDE zur Freigabe von Open Source Behörden-Software*

Seit Sommer 2022 bietet «Open CoDE» ([www.opencode.de](www.opencode.de)) eine Behörden-Plattform zur Freigabe von Open Source Software deutscher Verwaltungsstellen an (Tonekaboni, 2022). Diese Online-Plattform wurde durch das Bundesinnenministerium sowie die Bundesländer Baden-Württemberg und Nordrhein-Westfalen aufgebaut und fördert die Veröffentlichung und die Wiederverwendung von Behörden-Software in Deutschland. Neben der technischen, auf GitLab-basierten (Kelley, 2023) Open Source Plattform bietet «Open CoDE» auch Diskussionsforen, rechtliche Vorgaben, Lizenzrichtlinien, Code of Conduct und weitere Anleitungen und Empfehlungen an.

### *openDesk als digital souveräner IT-Arbeitsplatz*

Eines der auf «Open CoDE» intensiv entwickelten Open Source Projekte ist openDesk, der digital souveräne Arbeitsplatz für Mitarbeitende der öffentlichen Verwaltung in Deutschland. Initiiert hat das Projekt der IT-Rat im Oktober 2020 durch einen Auftrag an das Bundesministerium des Innern und für Heimat (BMI). Technisch besteht openDesk aus zahlreichen etablierten Open Source Applikationen, die nun eng miteinander verknüpft sind: eine Online-Office Lösung (Alternative zu Microsoft 365), Kalender und Email, eine Messenger-Lösung, Video Conferencing, Projektmanagement etc.

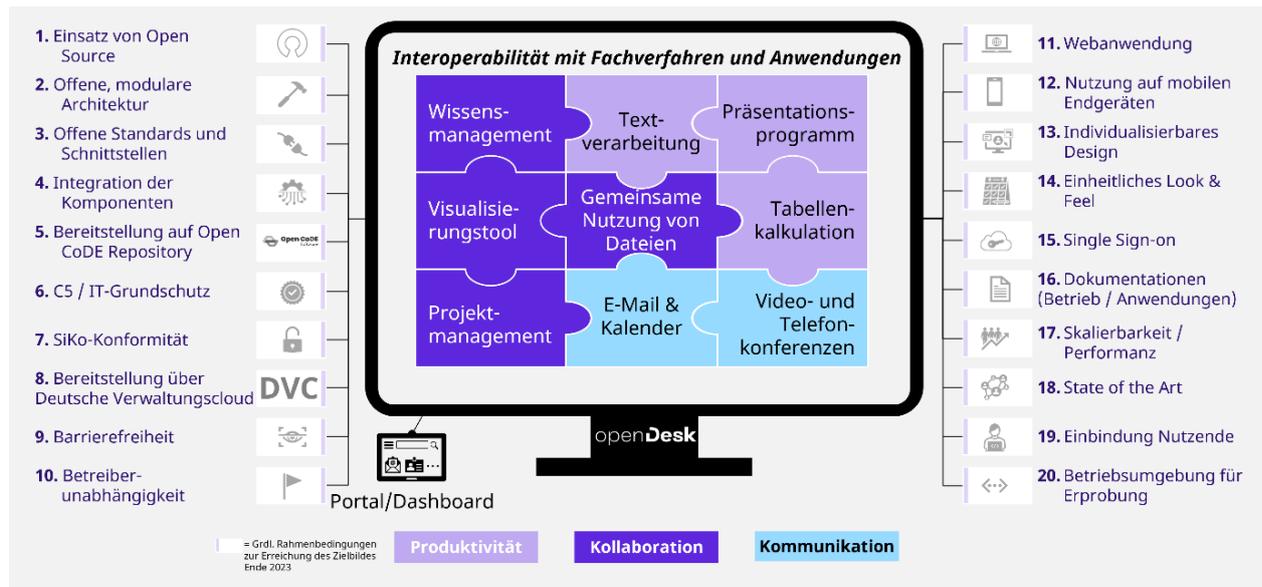

Abbildung 3: Produktvision von openDesk, dem digital souveränen Arbeitsplatz der deutschen Verwaltung (BMI, 2023)

*Sovereign Tech Fund*

Als übergeordnete, aber dennoch technologisch ausgerichtete Initiative gilt der «Sovereign Tech Fund». Über diese Initiative fördert die deutsche Regierung seit 2021 mit jährlich rund 10 Millionen Euro Open Source Software und Programmierbibliotheken, die an vielen Orten im öffentlichen und privaten Sektor eingesetzt werden (Groh et al., 2023). Diese Finanzierung verbessert die Sicherheit und Stabilität der digitalen Infrastruktur, die sowohl durch staatliche wie auch durch private Stellen intensiv genutzt wird (Pöting, 2022). Beispielsweise hat der «Sovereign Tech Fund» im November 2023 beschlossen, die Open Source Desktop-Umgebung GNOME mit einer Million Euro zu fördern, um die Barrierefreiheit zu verbessern (Tonekaboni, 2023).

### 3.2 Frankreich

Auch in Frankreich ist digitale Souveränität schon seit vielen Jahren ein Thema. Als einer der ersten hat der französische Radio-Pionier Pierre Bellanger bereits 2011 über digitale Souveränität als wichtiges Anliegen für die Wahrung der nationalen Interessen geschrieben (Bellanger, 2011). In seinem später veröffentlichten Buch «La souveraineté numérique» (Bellanger, 2014) kritisierte er die amerikanische Vorherrschaft im digitalen Raum und dass Frankreich und ganz Europa den US-Tech Firmen komplett ausgeliefert seien.

In diesem Sinne hat Frankreich acht Jahre später bei der Übernahme der EU-Präsidentschaft 2022 verkündet, dass sie die Thematik «Building Europe's Digital Sovereignty» aufgreifen wollen (Propp, 2022). Der im Zeitraum dieser Präsidentschaft erlassene «Digital Markets Act» zeigt exemplarisch auf, wie die digitale Souveränität auf regulatorischer Ebene in Europa gestärkt werden konnte (EURACTIV, 2022).

Als technische Massnahme hat die französische Regierung 2021 das Open Source Portal «Le Service Public du logiciel libre» code.gouv.fr lanciert, auf der über 9000 Software-Komponenten von über 100 Organisationen aus dem öffentlichen Sektor publiziert sind (Grzegorzewska, 2021). Ähnlich wie in Deutschland können so auch in Frankreich nationale, regionale und lokale Behörden eigene Software unter Open Source Lizenzen veröffentlichen, bestehende Anwendungen wiederverwenden und gemeinsam Programme weiterentwickeln.

Entsprechend aktiv haben sich in den letzten Jahren Tausende von französischen Open Source Anbietern und Nutzenden (unter anderem Behörden) in den Organisationen CNLL («Union des entreprises du logiciel libre et du numérique ouvert») und April («Association pour la Promotion et la Recherche en Informatique Libre») zusammengeschlossen, organisieren laufend Veranstaltungen und publizieren Berichte und Statistiken (Thévenet et al., 2023). Auch betreiben sowohl die französische Regierung als auch die Stadt Paris ein «Open Source Programme Office» (OSPO) zur Koordination aller Aktivitäten rund um Open Source Software (OpenForum Europe, 2022).

### 3.3 Europäische Kommission

Auch die Europäische Kommission ist digitalpolitisch aktiv. Gemäss der Schweizerischen Interdepartementalen Koordinationsgruppe EU-Digitalpolitik (IK-EUDP) des Bundes wurden zum Stichtag 15. März 2023 durch die EU insgesamt 35 Regulierungsmassnahmen und Initiativen bezüglich Digitalisierung in Europa identifiziert (IK-EUDP, 2023).

Mehrere Massnahmen davon werden durch die EU mit der Motivation für «digitale Souveränität» bzw. «technologische Souveränität» vorangetrieben. Übergeordnet steht das Konzept der «strategischen Souveränität», welche die EU wieder erlangen soll (Leonard and Shapiro, 2019). Die Zielsetzung der «digitalen Souveränität» wird durch die Europäische Kommission unter Ursula von der Leyen intensiv sichtbar gemacht (Burwell and Propp, 2020). Massgebend dabei ist das breite Verständnis, wie die EU ihre digitale Souveränität erhöhen könnte: Data Governance, Digital-Plattformen, digitale Infrastruktur, neue Technologien wie KI sowie Cybersecurity werden als wichtige Bereiche genannt (Roberts et al., 2021).

*Open Source Strategie der EU*

Die wohl älteste Aktivität zur Förderung der digitalen Souveränität in Europa ist die Unterstützung von Open Source Software. Schon früh wurde der hohe volkswirtschaftliche Nutzen von Open Source Software für Europa wissenschaftlich aufgezeigt (Ghosh, 2006). Diese Erkenntnisse bestätigt und erweitert auch ein neuerer, 390-seitiger Bericht des Fraunhofer-Instituts für System- und Innovationsforschung und von OpenForum Europe im Auftrag der Europäischen Kommission (Blind et al., 2021). Einerseits bewirkt die Entwicklung von Open Source Software in Europa jährlich einen Wertschöpfungszuwachs zwischen 65 und 95 Milliarden Euro (Schätzung für das Jahr 2018). Andererseits erhöht die Nutzung und Freigabe von Open Source Software die technologische Unabhängigkeit und stärkt damit die digitale Souveränität.

Diese positiven wirtschaftlichen und strategischen Effekte wurden durch die Europäische Kommission schon vor über 20 Jahren anerkannt. So fördert sie Open Source Software seit 2000 durch mehrere dezidierte Open Source Strategien, die sie in regelmässigen Abständen aktualisiert (EC, 2011). Die Strategie 2020 bis 2023 initiierte unter anderem die Gründung eines EU «Open Source Programme Office», das alle Tätigkeiten rund um Open Source Software koordiniert (OpenForum Europe, 2022).

Ausserdem betreibt die Europäische Kommission seit 15 Jahren das so genannte «Open Source Observatory» (OSOR), eine Informations- und Vernetzungsplattform bezüglich des Einsatzes und der Freigabe von Open Source Software in Europa. Mittels Länderberichten, Leitfäden und Empfehlungen (Devenyi et al., 2021), praxisnahen Fallstudien und einem Verzeichnis für Open Source Software hat sich OSOR zu einer internationalen und zuverlässigen Informationsquelle etabliert, die gemäss wissenschaftlicher Studien von hoher Qualität und Nutzen ist (Hollmann et al., 2013; Sowinska et al., 2021).

*GAIA-X*

In der europäischen IT-Branche das wohl bekannteste Vorhaben rund um digitale Souveränität ist GAIA-X, eine verteilte Cloud-Architektur zum Verarbeiten und Speichern von Daten in EU-Ländern (Braud et al., 2021). Erste Schritte wurden bereits 2015 von Deutschland und Frankreich unternommen, offiziell ist GAIA-X seit 2019 als offen zugängliche Cloud-Plattform gestartet (Fabbrini, Celeste and Quinn, 2020). Diese Offenheit, dass jegliche Firmen an GAIA-X teilnehmen können, wird inzwischen kritisch betrachtet und als mögliches «trojanisches Pferd» bezeichnet, da nun auch amerikanische Technologie-Konzerne wie Microsoft und auch chinesische Firmen mitwirken (Autolitano and Pawlowska, 2021).

*IPCEI Next Generation Cloud Infrastructure and Services*

Parallel zu GAIA-X (Krempl, 2021) treibt die Europäische Kommission im Rahmen des Programms «Important Project of Common European Interest» (IPCEI) das Vorhaben «Next Generation Cloud Infrastructure and Services» (CIS) voran (EC, 2023c). Das von Deutschland, Frankreich, Ungarn, Italien, der Niederlande, Polen und Spanien finanzierte Projekt wird mit 1.2 Milliarden Euro staatlich unterstützt und weitere 1.4 Milliarden Euro an Investitionen von europäischen Firmen generieren, wie im Dezember 2023 bekannt wurde (EC, 2023a). Mit diesem neuen Cloud-Ökosystem[2] sollen die technologischen Abhängigkeiten und damit die Lock-In Effekte reduziert werden, wie das Deutsche Ministerium für Wirtschaft und Klimaschutz ebenfalls im Dezember 2023 mitteilte (BMWK, 2023).

*EU Chips Act*

Ein aktuelles industriepolitisches Beispiel ist der EU Chips Act. Nach zwei Jahren Vorbereitung lancierte die Europäische Kommission im September 2023 unter dem Thema «Digital Sovereignty» den «European Chips Act». Dieser soll mit 3.3 Milliarden Euro die Forschung und Produktion von Halbleiterprodukten in Europa unterstützen und sieht zahlreiche weitere Massnahmen vor, wie die Halbleiterindustrie in Europa gestärkt werden soll (EC, 2023b).

*Weitere Regulierungen*

Neben den Förderinitiativen ist die EU im Kontext der digitalen Souveränität auch regulatorisch tätig. So schützt die 2018 in Kraft getretene Datenschutzgrundverordnung (DSGVO) die Daten der EU-Bevölkerung, was als direkter Beitrag zu einer Erhöhung der digitalen Souveränität angesehen wird (Gueham, 2017; Tambiama, 2020). Ausserdem soll der geplante «EU Data Act» die Interoperabilität der Daten zwischen den IT-Firmen verbessern. Dieser ermöglicht, dass Behörden, Unternehmen und Privatpersonen ihre eigenen Datenbestände einfach von einem Anbieter zu einem anderen wechseln können und so mehr digitale Souveränität und digitale Selbstbestimmung erlangen (Bühlmann and Reinle, 2022; European Commission, 2022).

Auch konnte im März 2024 mit dem «EU AI Act» das weltweit erste Gesetz zur Regulierung von künstlicher Intelligenz beschlossen werden (Europäisches Parlament, 2024). Damit werden unter anderem Open Source KI Modelle (siehe dazu Abschnitt «Künstliche Intelligenz») bevorzugt behandelt, da deren digitale Souveränität einen hohen Stellenwert hat und deshalb gefördert werden soll, wie jüngst ein Bericht in «Nature Briefing» erläuterte (Gibney, 2024). Zahlreiche weitere Regulierungen wie der Digital Services Act (DSA), der Digital Markets Act (DMA) and der Data Governance Act (DGA) spielen eine Rolle im Kontext von digitaler Souveränität, da sie den Willen der EU für eine klare Regulierung im digitalen Zeitalter markieren (Pons, 2023).

### 3.4 Indien

Digitale Souveränität ist in Indien ein heikles Thema, da oft ein direkter Bezug zur früheren Kolonialzeit hergestellt wird. So bezeichnen verschiedene indische Forschende die wachsende Abhängigkeit von ausländischen Technologiekonzernen als Neokolonialisierung, da Indien heute faktisch über Online-Plattformen und weitere Digitaltechnologien durch andere Länder kontrolliert wird (Gupta and Sony, 2021). Dieser sogenannte «Datenkolonialismus» umfasst unter anderem auch die Überwachung des Cyberspace (Couldry and Mejias, 2019b).

Als eine Antwort wird die lokale Datenverarbeitung («Data Localization») bezeichnet. So haben Forschende von Carnegie India einen umfassenden Bericht über die zahlreichen indischen Gesetze zur nationalen Kontrolle von Daten erarbeitet, der aufzeigt, wie ernsthaft dieses Anliegen von der indischen Regierung vorangetrieben wird (Burman and Sharma, 2021). Die Zielsetzung der digitalen Souveränität wird in Indien als neue Form der geopolitischen Ressourcenabsicherung betrachtet, der sogenannten «Data Securitisation» (Datenabsicherung) (Vila Seoane, 2021). Diese Perspektiven zeigen die strategische Relevanz der «Territorialität» von Daten auf und deuten darauf hin, mit welcher Wichtigkeit die lokale Datenverarbeitung und -speicherung in Indien verstanden wird.

Neben diesen übergeordneten Überlegungen haben die indische Regierung und die Wirtschaft auch erkannt, dass digitale Souveränität ein konkreter Weg darstellt, um die Abhängigkeit von ausländischen Technologiekonzernen zu reduzieren. So wurden in den letzten Jahren zahlreiche Digitaltechnologien gefördert, welche die Unabhängigkeit und Wahlfreiheit der indischen Gesellschaft erhöhten. Schon seit 2005 ist beispielsweise der Einsatz von Open Source Software durch die öffentliche Verwaltung obligatorisch (Avila Pinto, 2018). Und auch bei den Social Media Plattformen wurde die Abhängigkeit von ausländischen Technologiekonzernen reduziert. So hat «Koo», die indische Alternative zu Twitter (bzw. X), im Jahr 2021 einen Aufschwung erlebt und auch Millionen von Downloads aus Brasilien generiert (Chandran, 2023). Weiter berichtet Thomson Reuters, dass Plattformen wie «MapmyIndia» eine relevante Konkurrenz zu Google Maps darstellen und das indische Mobile Betriebssystem BharOS eine Alternative zu Google Android bietet (Chandran, 2023). Neben amerikanischen Konzernen wehrt sich Indien auch gegen Technologieunternehmen aus China, die mit Online-Plattformen wie TikTok und der Telekommunikations-Hardware von Huawei ihre technologische Dominanz ausüben wollen.

---

[2] https://www.ipcei-cis.eu

### 3.5 Lateinamerika

Lateinamerikanische Länder wie Brasilien, Venezuela, Ecuador, Bolivien und Uruguay haben eine jahrzehntelange Tradition der digitalen Souveränität. Bereits in den frühen 2000er-Jahren haben diese Länder Gesetze erlassen, welche die staatliche Datenverarbeitung mittels Open Source Software vorschreiben (Avila Pinto, 2018). Insbesondere Kuba, welches das US-amerikanische Software-Embargo umgehen musste, war auf eine Alternative zu Microsoft Windows angewiesen. So führte die kubanische Regierung 2009 die Open Source basierte Linux-Distribution «Nova» in allen Behörden ein, sparte damit hohe Lizenzkosten und war gleichzeitig weniger verwundbar gegenüber Malware und versteckten Hintertüren von Microsoft (Henken and Garcia Santamaria, 2021).

Dennoch stellen einige Forschende fest, dass die Abhängigkeit lateinamerikanischer Länder von US-amerikanischen und chinesischen Technologiekonzernen weiterhin hoch ist (Becerra and Waisbord, 2021). Es wird gar von einer «laissez-faire» Deregulierung gesprochen, welche ausländische IT-Unternehmen einlädt, sich in den lateinamerikanischen Ländern niederzulassen und damit die Abhängigkeiten noch erhöht.

Aus geopolitischer Sicht ist auch Lateinamerika von einem digitalen Neokolonialismus betroffen, der, wie erwähnt, als «Datenkolonialismus» bezeichnet wird (Couldry and Mejias, 2019b). Diesbezüglich zeigen Forschende auf, dass die Verfestigung dieser Ungleichheiten in der digitalen Welt von lateinamerikanischen Ländern noch zugenommen hat (Tait, dos Reis Peron and Suárez, 2022). Insgesamt wird auch von «Digitalkolonialismus» gesprochen, der primär aus den USA auf den globalen Süden über Technologie-Abhängigkeiten und Überwachungsmöglichkeiten ausgeübt wird (Kwet, 2019). Im Zentrum der Kritik stehen dabei die Suchmaschine, der Web-Browser Chrome und das Smartphone-Betriebssystem Android von Google, die Cloud-Services und das Desktop-Betriebssystem Windows von Microsoft, Social Media Plattformen von Facebook (Meta) und Twitter (jetzt X), Video-Streaming von Netflix, Youtube und weitere Online-Angebote von US-amerikanischen Technologieunternehmen.

### 3.6 USA

Obwohl die USA mit Blick auf die grossen, US-amerikanischen Technologie-Konzernen zu den Gewinnern der Digitalisierung gehören, gibt es insbesondere in Bezug auf China auch wichtige Abgrenzungsbedürfnisse. So spielten Überlegungen bezüglich digitaler Souveränität in den letzten Jahren eine zentrale Rolle beim Exportverbot von Software an chinesische Unternehmen wie beispielsweise Googles Mobile-Betriebssystem Android für die Smartphones von Huawei (Cartwright, 2020). Auch Online-Plattformen wie Tiktok der Firma ByteDance oder WeChat von Tencent sowie der mächtige Cloud-Anbieter Alibaba wurden durch die amerikanische Politik kritisiert und teilweise eingeschränkt. Diese Beispiele zeigen, wie selbst die grossen Staaten ihre digitale Souveränität gegenüber der digitalen Expansion von anderen Ländern schützen wollen, um mehr Kontrolle über Digitaltechnologien zu erlangen (Roberts, Hine and Floridi, 2023).

### 3.7 China

Umgekehrt ergeht es China, das sich von den USA insbesondere im digitalen Raum möglichst unabhängig positionieren will. Bereits 2011 wurde der Begriff «Cyber Sovereignty» im Kontext der unabhängigen Kontrolle der nationalen Internet-Domains verwendet, sodass China auch heute noch seine «Great Firewall» uneingeschränkt durchsetzen kann. In 2016 haben die zwei chinesischen und russischen Staatspräsidenten Vladimir Putin und Xi Jinping eine Vereinbarung unterzeichnet, in der sie die Wichtigkeit der digitalen Souveränität betonten und sich gegenseitig Unterstützung zusicherten (Zinovieva and Shitkov, 2023). In all den Jahren konnte China ausserdem seine Zensurregeln den US-amerikanischen Firmen wie Google und Meta aufzwingen, sodass das Land eine vollständige Kontrolle über die ausländischen Internet-Dienste behalten konnte. So setzt China digitale Souveränität auch primär mit dem Ziel ein, die vollständige Kontrolle über den digitalen Raum auszuüben und so den Einfluss der Regierung im In- und Ausland zu maximieren (McKune and Ahmed, 2018). Dennoch bleibt die radikal implementierte digitale Souveränität in China ein schwieriges Thema, da deren autoritär realisierte Umsetzung im Widerspruch zum globalen Internet steht (Hong and Goodnight, 2020).

### 3.8 Internationales Komitee vom Roten Kreuz (IKRK)

Das Internationale Komitee vom Roten Kreuz (IKRK) ist im Gegensatz zu den einzelnen Ländern angewiesen, eine möglichst neutrale Position einnehmen zu können, um in Konfliktsituation das Vertrauen aller Parteien zu geniessen. Insbesondere im digitalen Raum ist es jedoch für das IKRK zunehmend schwierig, unabhängig von Technologie-Unternehmen und deren Ursprungsländer Daten zu speichern und zu verarbeiten (ICRC, 2022). Die Anforderungen sind hoch, denn es fallen viele hochsensible Daten an, die mit verschiedensten Akteuren ausgetauscht werden müssen – «digitale Immunität» ist gefordert (Le Blond *et al.*, 2018). Darum machen sich Führungspersonen Gedanken, wie das IKRK seine digitale Souveränität und damit die Glaubwürdigkeit der gesamten Organisation im digitalen Zeitalter bewahren kann. Um die Cyberrisken zu reduzieren, Abhängigkeiten von IT-Firmen zu reduzieren und die Kontrolle über die Digitaltechnologien wieder zu erlangen, prüft das IKRK den verstärkten Einsatz von Open Source Software (Devidal, 2023).

### 3.9 Vereinte Nationen

Die Vereinten Nationen haben sich bisher nicht explizit zum Thema digitale Souveränität geäussert, die grundlegenden Ideen dahinter werden im internationalen Kontext jedoch schon seit längerem diskutiert (Zinovieva and Shitkov, 2023). So wurden im Rahmen von UN-Berichten von zwei Arbeitsgruppen in den letzten Jahren konkrete Empfehlungen bezüglich ICT-Einsatz und Souveränität vermittelt. Bereits 2015 hat die «Group of Governmental Experts» auf die Gefahr der kriminellen Verwendung von ICT hingewiesen und die Staaten aufgerufen, entsprechende Massnahmen zur Absicherung ihrer digitalen Infrastruktur vorzunehmen (GGE, 2015). Die «Open-ended Working Group» (OEWG) hat 2021 im Schlussbericht über die ICT-Entwicklungen im Kontext der internationalen Sicherheit festgehalten, dass Digitaltechnologien beim Einsatz gegen kritische Infrastrukturen die staatliche Souveränität verletzen können (OEWG, 2021). Im Oktober 2023 wurde im UN-affiliierten «Internet Governance Forum» (IGF) intensiv über digitale Souveränität und Nachhaltigkeit debattiert (IGF, 2023). Und auch in der Forschung wird auf die Notwendigkeit der Diskussion zu digitaler Souveränität in multilateralen Gremien wie den Vereinten Nationen hingewiesen, insbesondere bezüglich den geopolitischen Auswirkungen der Regulierung von künstlicher Intelligenz (Roberts, Hine and Floridi, 2023).

*High-Level Panel on Digital Cooperation*

Die kritische Rolle von Technologieunternehmen in Bezug auf digitale Souveränität wurde in den Vereinten Nationen bisher jedoch kaum thematisiert. Das kann auch damit zusammenhängen, dass diese Firmen typischerweise seit jeher gute Beziehungen zu UN-Organisationen pflegen, aktiv in Arbeitsgremien mitwirken und dadurch entsprechende Kritik vorbeugen können.

So wurde beispielsweise die offizielle Agenda der Vereinten Nationen bezüglich digitaler Transformation unter der Leitung von Melinda Gates von der «Bill & Melinda Gates Foundation» und Jack Ma als Vorstandsvorsitzender der chinesischen Alibaba Group im Rahmen eines «High-level Panel on Digital Cooperation» im Auftrag des UN-Generalsekretärs lanciert (Digital Cooperation, 2019). Deshalb erstaunt es nicht, dass in diesem Bericht nicht die Abhängigkeit von grossen IT-Unternehmen oder der Datenkolonialismus durch Technologie-Firmen kritisiert wird. Vielmehr geht es um die Frage, wie beispielsweise der globale «Digital Divide» reduziert und die Nachhaltigkeits-Agenda («Sustainable Development Goals») durch mehr Bildung zu Digitaltechnologien und Kooperation im digitalen Raum erreicht werden kann.

Dennoch hat diese vielbeachtete und hochrangig besetzte Multi-Stakeholder Initiative (Prestes, 2019) indirekt auch einen Beitrag zur Förderung der digitalen Souveränität durch die Vereinten Nationen geleistet. So werden im Report erfolgreiche Open Source Projekte wie die «Modular Open Source Identity Platform» (MOSIP) vorgestellt, die ein Land zur elektronischen Identifikation verwenden kann. Eine der Empfehlungen des «High-level Panel on Digital Cooperation» schaffte ausserdem die Grundlage der sogenannten «Digital Public Goods Alliance» (United Nations General Assembly, 2020).

*Digital Public Goods*

Das von den Vereinten Nationen initiierte Konzept der «Digital Public Goods» basiert auf dem Gedanken, dass einmal erstellte digitale Güter wie Software, Daten, digitale Inhalte und Modelle der künstlichen Intelligenz beliebig oft verwendet werden können, ohne dass sie an Leistung oder Qualität verlieren. Entscheidend dabei ist, dass die Urheberrechte die uneingeschränkte Nutzung, das Kopieren und die Weiterentwicklung erlauben (Digital Public Goods Alliance, 2021).

Wenn digitale Güter somit den Charakter von öffentlichen Gütern erhalten und damit weder rivalisierend noch ausschliessbar sind (Samuelson, 1954), werden digitale Güter zu «digitalen öffentlichen Gütern». Dies ist grundsätzlich der Fall bei Open Source Software, Open Data, Open Standards, offenen Modellen der künstlichen Intelligenz («Open AI models») und bei Open Content (Digital Public Goods Alliance, 2021). Gemäss der 2020 veröffentlichten «Roadmap for Digital Cooperation» sollen die sogenannten «Digital Public Goods» aber auch zur Erreichung der Nachhaltigkeitsziele beitragen und weitere Charakteristiken aufweisen, welche sie langfristig nutzbar machen (UN Secretary General, 2020).

Dazu hat die «Digital Public Goods Alliance» einen Standard mit detaillierten Kriterien entwickelt, die erfüllt sein müssen, damit ein Open Source Projekt oder eine Open Data Sammlung als «Digital Public Good» bezeichnet werden kann. Dieser «Digital Public Goods Standard» (Digital Public Goods Alliance, 2020) umfasst folgende neun Vorgaben, die eine hohe Ähnlichkeit mit den Voraussetzungen für digitale Nachhaltigkeit (Stürmer, Abu-Tayeh and Myrach, 2017; Stürmer *et al.*, 2023) aufweisen:

1. *Relevanz für die Nachhaltigkeitsziele:* Digitale öffentliche Güter müssen zu konkreten Nachhaltigkeitszielen («Sustainable Development Goals» SDGs) beitragen.
2. *Offene Lizenz:* Software muss unter einer Open Source Lizenz der Open Source Initiative (OSI) veröffentlicht sein. Daten, Inhalte und KI-Modelle müssen unter Creative Commons Lizenzbedingungen freigegeben sein.
3. *Klarheit beim geistigen Eigentum:* Die Urheberrechte müssen deklariert und alle rechtlichen Bedingungen bezüglich geistiges Eigentum müssen eingehalten werden.
4. *Plattform-Unabhängigkeit:* Digitale öffentliche Güter dürfen keine Abhängigkeiten von proprietären Technologien enthalten.
5. *Umfassende Dokumentation:* Bei Software und KI-Modellen muss eine umfassende technische Spezifikation vorhanden sein. Datensätze müssen gut dokumentierte Metadaten beinhalten.
6. *Export-Möglichkeit der Daten:* Bei digitalen öffentlichen Gütern müssen alle Daten der Nutzenden in einem offenen Standard importierbar und exportierbar sein.
7. *Einhaltung der Gesetze:* Digitale öffentliche Güter müssen so konzipiert sein, dass sie den Datenschutz und auch allen anderen geltenden Gesetzen (bspw. EU AI Act, EU Digital Services Act, EU Digital Markets Act etc.) einhalten.
8. *Umsetzung von Standards und Best Practices:* Digitale öffentliche Güter müssen so gestaltet sein, dass sie etablierte Grundsätze für Usability, Skalierbarkeit, Sicherheit etc. erfüllen.
9. *«Do No Harm by Design»:* Digitale öffentliche Güter dürfen von ihrer Zielsetzung her keinen Schaden anrichten und müssen folgende Aspekte einhalten:
    a) *Datenschutz und Datensicherheit:* Digitale öffentliche Güter, die personenbezogene Daten beinhalten, müssen die Privatsphäre schützen und die Sicherheit der Daten gewährleisten.
    b) *Unangemessene und illegale Inhalte:* Digitale öffentliche Güter, die Inhalte von Nutzenden sammeln und verbreiten, müssen ermöglichen, dass unangemessene und illegale Inhalte identifiziert und gemeldet werden können.
    c) *Schutz vor Belästigung:* Digitale öffentliche Güter müssen über Möglichkeiten verfügen, Minderjährige vor Missbrauch zu schützen.

Die «Digital Public Goods Alliance" hat eine Plattform zur Verfügung gestellt, auf der Software, Datensammlungen, digitale Inhalte, KI-Modelle etc. eingereicht werden können, um sie in die sogenannte «Registry» von digitalen öffentlichen Gütern aufzunehmen. Aktuell sind 171 digitale öffentliche Güter registriert, die Mehrheit davon (142) sind Open Source Projekte (siehe Abbildung 3).

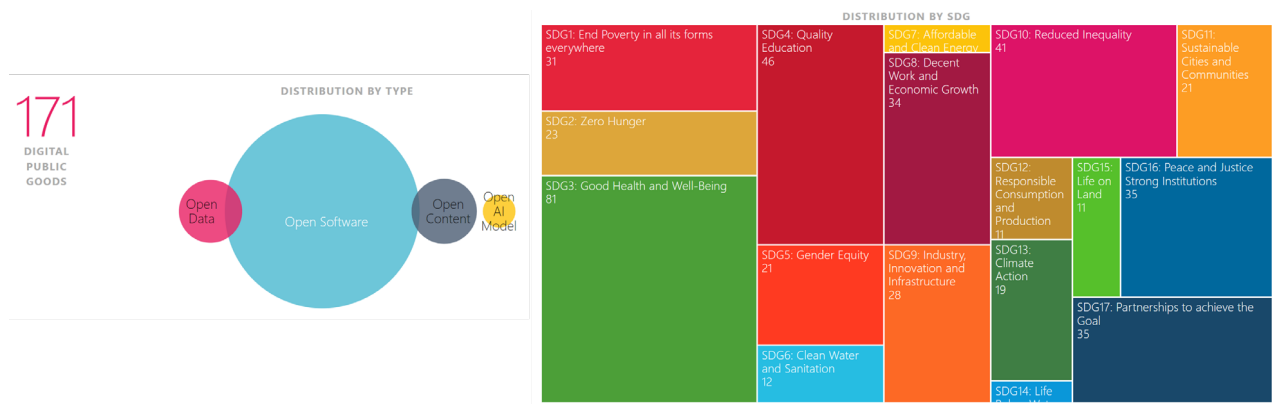

Abbildung 4: Registrierte digitale öffentliche Güter auf der Online-Plattform der «Digital Public Goods Alliance»
https://digitalpublicgoods.net/registry/

Bezüglich digitaler Souveränität zeigt der Einsatz und die Freigabe von digital öffentlichen Gütern konkrete Lösungswege auf, wie Staaten ihre Abhängigkeit gegenüber IT-Herstellern reduzieren und die Kontrolle über die digitale Infrastruktur erhöhen können (Nordhaug and Harris, 2021). Beispielsweise konnte die Open Source Gesundheits-Management Software «District Health Information Software version 2» (DHIS2) als digitales öffentliches Gut in zahlreichen Ländern für die bessere Gesundheitsversorgung eingesetzt werden (Nicholson *et al.*, 2022). So können die Länder nun selbständig den Betrieb wählen und gemeinsam über die Open Source Community die Software weiterentwickeln, was die digitale Souveränität konkret fördert.

*Indigenous Data Sovereignty*

Aus internationaler Sicht bemerkenswert ist auch die Bewegung zu «Indigenous Data Sovereignty» (Kukutai, 2023). Forschende mit indigenem Hintergrund aus Neuseeland haben 2016 eine Agenda zur Datensouveränität von indigenen Menschen entwickelt (Taylor and Kukutai, 2016). Basierend auf der «United Nations Declaration on the Rights of Indigenous Peoples» (UNDRIP) wird auf die Wichtigkeit der Souveränität im digitalen Raum für Indigene hingewiesen. Beispielsweise zeigen die 5D-Begriffe «disparity, deprivation, disadvantage, dysfunction and difference» auf, wie Daten über Aborigines die «Ungleichheit, Entbehrung, Benachteiligung, Dysfunktion und Differenz» mit der nicht-indigenen Australischen Bevölkerung messen (Walter, 2016). Ein weiteres Buch zeigt die Notwendigkeit von «Indigenous Data Governance» auf, also wie indigene Völker weltweit Daten benötigen um ihre Rechte wahrnehmen zu können (Walter et al., 2021). Konkret geht es auch darum, dass indigene Menschen Zugang zu Technologie und Open Government Data erhalten und so die Kontrolle über ihre Daten wiedererlangen (Diviacchi, 2023; Hudson et al., 2023). Insbesondere bei der Nutzung von frei zugänglichen Daten ist es gemäss «Indigenous Data Sovereignty» Forschenden wichtig, die so genannten CARE Prinzipien «Collective Benefit, Authority to Control, Responsibility, and Ethics» (kollektiver Nutzen, Vollmacht der Kontrolle, Verantwortung und Ethik) einzuhalten um die Kontrolle über die Daten wiederzuerlangen (Carroll et al., 2020; Leonard et al., 2023).

# 4 Strategische Stossrichtungen

Zur Erarbeitung der Strategie «Digitale Souveränität der Schweiz» wird empfohlen, Stossrichtungen in zwei Dimensionen zu berücksichtigen: Einerseits betreffen dies die unterschiedlichen Themenfelder Software, Daten, IT-Infrastruktur und künstliche Intelligenz. Andererseits sind auch Sektor-spezifische Aspekte aus den unterschiedlichen Branchen zu berücksichtigen. Einen Gesamtüberblick schafft Tabelle 2, die alle in diesem Bericht erläuterten Aktivitäten zu digitaler Souveränität in der Schweiz und im Ausland nach den verschiedenen Sektoren und Themenfeldern auflistet. Im Gebiet der künstlichen Intelligenz konnten bis anhin noch keine Sektor-spezifischen Aktivitäten festgestellt werden:

| *Themenfeld:*<br>Sektoren: | Software | Daten | IT-Infrastruktur | Künstliche Intelligenz |
|---|---|---|---|---|
| **Finanzen** | • Open Banking | • EU Digital Identity Wallet | • Digital Operational Resilience Act | ? |
| **Gesundheit** | • Digisanté | • Digisanté<br>• Swiss Personalized Health Network<br>• European Health Data Space (EHDS) | • Digisanté | ? |
| **Verkehr** | • MODIG<br>• Open Journey Planner (OJP) | • MODIG<br>• NADIM<br>• EU Mobility Data Space | • MODIG | ? |
| **Bildung** | • dBildungscloud | • Edulog<br>• Switch edu-ID<br>• dBildungscloud<br>• Data Space Education and Skills (DASES) | • Open Education Server | ? |

Tabelle 2: Überblick Aktivitäten zu digitaler Souveränität in unterschiedlichen Themenfeldern und Sektoren

### 4.1 Technologische Themenfelder

Digitaltechnologien können in verschiedene Themenfelder unterteilt werden. In diesem Bericht wird nach Software, Daten, IT-Infrastruktur und künstlicher Intelligenz unterschieden.

*Software*

Durch das Schreiben von Quellcode in unterschiedlichsten Programmiersprachen wird Software entwickelt. Software kann auf Smartphones und Tablets, auf Laptops oder Desktop Computer (Clients), auf Servern und eingebettet in unterschiedlichste Umgebungen (Fernsehen, Auto, Heizungen etc.) betrieben werden. Auch Teil von Software sind Schnittstellen, so genannte «Application Programming Interfaces» (APIs), die für den Datentransfer zwischen interoperablen IT-Systemen verwendet werden. Software bildet damit einen fundamentalen «Building Block» der digitalen Infrastruktur (Digital Public Goods Alliance, 2022).

Digitale Souveränität auf der Ebene von Software zeichnet sich einerseits dadurch aus, dass die Software eigenständig entwickelt, verändert, kontrolliert und ergänzt werden kann (siehe Seite 6, Abschnitt 1.1 «Definition digitale Souveränität»). Voraussetzung dafür ist der technische Zugang zum Quellcode und die rechtliche Möglichkeit, diesen zu verändern und uneingeschränkt einsetzen zu können. Dies entspricht der Open Source Definition[3] der Open Source Initiative, die bei der Aufnahme von Open Source Lizenzen angewendet wird (bitkom, 2023).

Andererseits ist bezüglich digitaler Souveränität von Software auch das Format für die Datenspeicherung von Bedeutung. Verwenden Applikationen proprietäre Datenformate, können bspw. Entsprechende Dokumente nur durch diese Anwendungen geöffnet und verarbeitet werden, was die System-Abhängigkeit unnötig erhöht. Entscheidend ist deshalb, dass die Daten in offenen Standards gespeichert werden, sodass die Daten auch durch andere Programme weiterverarbeitet werden können. Ein offener Standard definiert sich über folgende fünf Eigenschaften (Tiemann, 2006): Verbreiteter Einsatz des Standards, internationale Akzeptanz (nicht nur nationaler Standard), Weiterentwicklung durch konsensbasierte Standardisierungs-Organisation (bspw. ISO, W3C, OASIS), Standard verlangt keine Lizenzgebühren und enthält keine Patente sowie der Standard ist klar und offen dokumentiert über eine fundierte Spezifikation. Sind diese Voraussetzungen erfüllt und wird der offene Standard ausserdem über eine Software-Bibliothek unter einer Open Source Lizenz implementiert, dann ist ein Höchstmass an digitaler Souveränität gewährleistet.

*Daten*

Daten umfassen alle Inhalte wie Texte, Bilder, Zahlen etc., die in unterschiedlichen Formaten (Datenstandards) gespeichert werden. Darunter fallen auch Identitätsdienste wie bspw. die staatliche E-ID und das dazugehörige Prinzip «Self-Sovereign Identity» (Bundesrat, 2023c). Daten werden von Behörden, Unternehmen und Individuen produziert. Beispiele wie OpenStreetMap oder Wikipedia zeigen auf, dass auch zivilgesellschaftlich gesammelte Daten verlässlich sind (Goodchild and Li, 2012) und deshalb produktiv eingesetzt werden können (Hitz-Gamper and Stürmer, 2021). Wenn die Daten personenbezogene Informationen enthalten, fallen sie unter das Datenschutzrecht, wenn es sich um Sachdaten handelt, dann gilt der Datenschutz nicht. Auch Daten können unter offenen Lizenzen freigegeben werden, wozu seit über zwanzig Jahren meist Creative Commons Lizenzen angewendet werden (Lessig, 2004).

---

[3] https://opensource.org/osd/

Im Kontext von digitaler Souveränität wird bei Daten gemäss obenstehender Definition der Anspruch erhoben, dass uneingeschränkte Kontrolle über die Datenspeicherung und Datenverarbeitung besteht und dass ausserdem eigenständig bestimmt werden darf, wer Zugriff auf die Daten hat. Können Behörden somit nicht ausschliessen, dass auch ausländische Staaten oder unbefugte Firmen Zugang zu staatlichen Daten haben, kann nicht von digitaler Souveränität oder Datensouveränität gesprochen werden. Insofern erscheint es fraglich, wenn Firmen wie Microsoft oder SAP behaupten, dass sie durch ihre Public Cloud Dienste die digitale Souveränität verbessern wollen (Innovate Switzerland, 2023).

*IT-Infrastruktur*

Mit IT-Infrastruktur werden sämtliche Hardware-Bestandteile der Informatik und Telekommunikation sowie den benötigten Betriebssystemen verstanden. So betrifft dies einerseits die physische Welt der Digitaltechnologien, die mit Clients und Servern sowie deren Bestandteile wie «Central Processing Unit» (CPU) und «Graphics Processing Unit» (GPU) etc. die eigentlichen Rechenkapazitäten zur Verfügung stellen. Andererseits werden mit dem Betriebssystem auch Software-Elemente dazugezählt, welche die Grundlage für das Funktionieren von Applikationen und Middleware (Datenbanken, Container-Management etc.) bilden.

In Bezug auf digitale Souveränität ist die Verfügbarkeit und der Zugriff auf die IT-Infrastruktur entscheidend. Bei Cloud-Computing Diensten betrifft dies die Server-Hardware, deren Betrieb in Rechenzentren sowie die gesamte Telekommunikationsinfrastruktur für die Datenübermittlung. Gemäss dem Schichtenmodell von acatech (Kagermann, Streibich, and Suder, 2021) werden dazu alle benötigten Rohmaterialien (Seltene Erden etc.) und Vorprodukte (Batterien etc.), Komponenten (Mikrochips wie CPUs und GPUs etc.) und die Kommunikationsinfrastruktur (Mobilfunknetze, Galileo-Navigation etc.) benötigt. Zusammengefasst kann diese unterste Technologieschicht als «Infrastructure-as-a-Services» bezeichnet werden.

In Bezug auf digitale Souveränität stellt sich dabei die Frage, welche Rolle die Verfügbarkeit von IT-Hardware auf den unteren Technologieschichten spielt. Obwohl IT-Infrastruktur als Commodity betrachtet werden kann, da sie bloss eine geringe Hersteller-Abhängigkeit verursacht, kann die Knappheit der physischen Ressourcen und die Wirtschaftlichkeit des Betriebs die digitale Souveränität letztlich doch beeinträchtigen. Auch haben Hardware-Anbieter theoretisch die Möglichkeit, sogenannte «Backdoors» in IT-Komponenten einzubauen um später auf Systeme und Daten zugreifen zu können (Hashemi and Zarei, 2021). So ist es von Vorteil, nicht von einzelnen Hardware-Lieferanten abhängig zu sein, weshalb auch die Entwicklung und Herstellung von Halbleitererzeugnissen Sinn macht (siehe Seite 16 «EU Chips Act»). Ausserdem haben «Hyperscalers» mit ihren Public Cloud Services eine hohe Anzahl gut ausgebildeter Mitarbeitende, was ihnen deutliche Marktvorteile erbringt (siehe dazu auch Seite 9, Abschnitt 1.4 «Trade-off zwischen Public Cloud Services und digitaler Souveränität»). Insofern muss der Staat bereit sein, für eigenständig betriebene IT-Infrastruktur auch höhere finanzielle Auslagen zu tätigen (siehe dazu auch Seite 32 «Massnahme 9: Aufbau der Swiss Government Cloud basierend auf Open Source Technologien»).

*Künstliche Intelligenz*

Auch wenn künstliche Intelligenz (KI) auf Software, Daten und IT-Infrastruktur basiert, so stellt diese Technologie aufgrund der Verflechtung der zuvor beschriebenen Elemente und ihrer einzigartigen Auswirkungen ein Themenfeld dar, das einer gesonderten Betrachtung bedarf. Beispielhaft zeigt sich dies an der Anzahl Forschungspublikationen zu KI, die sich von 200'000 Veröffentlichungen im Jahr 2010 auf fast 500'000 Studien in 2021 mehr als verdoppelte (Maslej *et al.*, 2023) und in den Folgejahren wohl nochmal stark zugenommen hat. Dabei wächst der Einfluss des Privatsektors deutlich: Währenddem im Jahr 2006 die grössten KI-Modelle noch vollständig aus der Wissenschaft stammten, kamen 2021 nun 96% der grossen KI-Modelle von IT-Firmen (Ahmed, Wahed and Thompson, 2023). Dies zeigt die rasche Privatisierung der KI-Technologien und die Dominanz der IT-Industrie in der KI-Forschung.

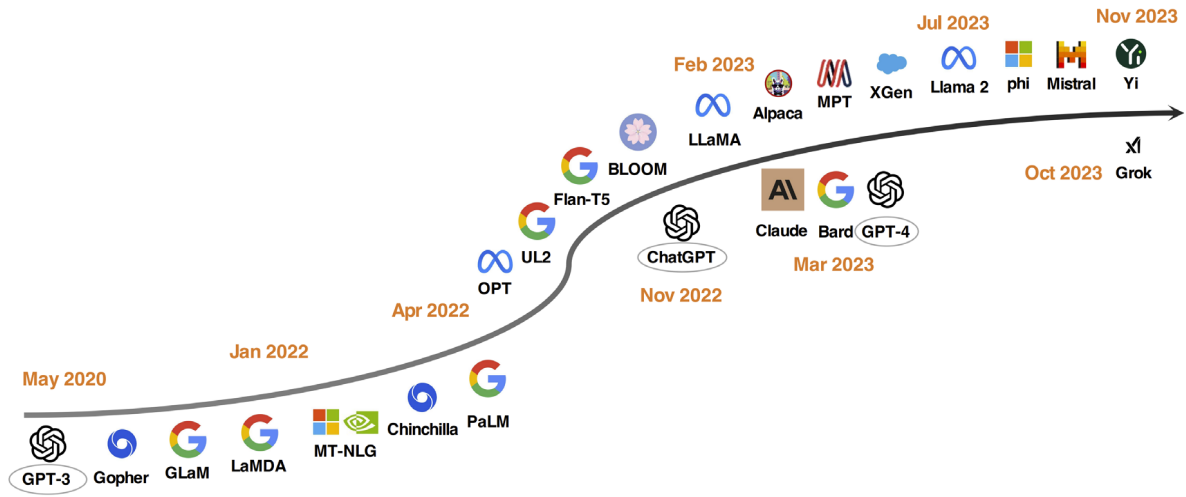

Abbildung 5: Entstehung von «Large Language Models» (LLMs) in den letzten vier Jahren. Die Modelle unterhalb des Pfeils sind proprietär, die Modelle oberhalb gelten als «Open Source KI Modelle» (Chen *et al.*, 2023)

Umso wichtiger ist es in Bezug auf digitaler Souveränität, dass die Entwicklung und die Nutzung von KI-Modellen wieder vermehrt in die Kontrolle des öffentlichen Sektors gelangt. So raten auch die Vereinten Nationen in ihrem Zwischenbericht «Governing AI for Humanity» dazu, dass sogenannte «Open Source KI Modelle» («Open Source AI Models») und die dazu notwendigen Daten und Hyperparameter (Einstellungen) gefördert werden um Transparenz, Vertrauen und Rechenschaft zu verbessern (United Nations, 2023). Diesbezüglich findet zurzeit ein Wettlauf zwischen proprietären Modellen wie GPT der Firma OpenAI (OpenAI *et al.*, 2023) und Open Source KI Modellen wie BLOOM (Le Scao *et al.*, 2022) oder Mistral/Mixtral (Jiang *et al.*, 2024) statt, bei dem die offenen Modelle immer näher an die Leistung von den geschlossenen Modellen gelangen (Chen *et al.*, 2023). Auch setzen einzelne Länder wie bspw. Dänemark gezielt auf die Entwicklung von nationalen «Foundation Models», also KI-Modelle, welche die lokale Sprache und Begrifflichkeiten verstehen (Enevoldsen *et al.*, 2023).

### 4.2 Sektorspezifische Stossrichtungen

Nach der technologischen Vertiefung werden im Folgenden spezifische Sektoren betrachtet, die exponiert sind und bei denen deshalb die Erhöhung der digitalen Souveränität besonders wichtig ist. Beispiele aus den betrachteten Sektoren zeigen auf, weshalb gerade dort digitale Souveränität ein zentraler strategischer Wert darstellt. Auch wird jeweils auf internationale Entwicklungen hingewiesen, die für die Schweiz von Bedeutung sind.

*Finanzwesen*

Der Finanzmarkt gilt als einer der zentralen Schweizer Wirtschaftszweige und auch die heimische Wirtschaft benötigt verlässliche Finanzdienstleistungen. So gelten diese gemäss der «Nationalen Strategie zum Schutz kritischer Infrastrukturen» als Infrastruktur mit sehr hoher Kritikalität (Bundesrat, 2023d). Um täglich Millionen von Geldtransaktionen ausführen zu können, werden stabile Kommunikationsinfrastrukturen und zuverlässige Digitaltechnologien benötigt. Dies weist auf den hohen Stellenwert der digitalen Souveränität im Finanzsektor hin, da sowohl die Banken und Versicherungen als auch alle anderen Schweizer Wirtschaftsakteure unabhängig von ausländischen Technologieunternehmen ihre Finanztransaktionen ausführen wollen.

Gleichzeitig ist die Finanzbranche international ausgelegt, weshalb eine Abschottung des nationalen Zahlungsverkehrs wenig Sinn ergeben würde. Dies deutet auf die besondere Herausforderung hin, dass sowohl der internationale Datenaustausch als auch die nationale Souveränität stets gewährleistet sein müssen. Wie problematisch eine Abkopplung aus dem internationalen Zahlungsverkehr sein kann, zeigen jüngste Forschungsergebnisse bezüglich der wirksamen Sanktionen gegenüber russischen Banken durch den Ausschluss vom SWIFT-System der «Society for Worldwide Interbank Financial Telecommunication» (Drott, Goldbach and Nitsch, 2024).

Ein weitere Aspekt betrifft die Währungssouveränität eines Landes, die durch Kryptowährungen bedroht werden. So hat die EU im sogenannten «Digital Finance Package» mehrere Regularien erlassen, um die digitale Souveränität auch im Crypto Finance Umfeld zu gewährleisten (Donnelly, Ríos Camacho and Heidebrecht, 2023). Als weitere EU-Regulierung will der «Digital Operational Resilience Act» (DORA)[4] die Stabilität und Verfügbarkeit der IT-Systeme für Banken und andere Finanzdienstleister sicherstellen und sie gegen Cyberattacken schützen. Damit stärkt die 2023 in Kraft getretene DORA-Regulierung indirekt auch die digitale Souveränität, wie Forschende in einer umfassenden Studie festhalten (Donnelly, Ríos Camacho and Heidebrecht, 2023).

Im Zusammenhang mit Finanzregulierung auch wichtig ist die eindeutige Identifikation von natürlichen und juristischen Personen («Know Your Customer»), damit Bankkonten eröffnet und Finanztransaktionen ausgeführt werden können. Dies wird in der Europäischen Union mittels der «Electronic Identification, Authentication and Trust Services» (eIDAS)[5] Regulierung und dem «EU Digital Identity Wallet» (EU-DIW)[6] sichergestellt. Diese digitale Brieftasche gilt als weiterer Schritt der EU in Richtung digitale Souveränität, weil die Identifikation unabhängig von Technologieunternehmen geschieht. Im Gegenteil, Firmen wie Amazon oder Booking.com müssen künftig den EU-Standard selber implementieren um in der EU weiterhin kommerziell tätig zu sein (European Commission, 2023).

Um die Zusammenarbeit zwischen den Banken und anderen Finanzinstituten zu erleichtern und damit die digitale Souveränität der einzelnen Akteure zu erhöhen, werden im Finanzsektor oft «Open Banking» Prinzipien angewendet (Long *et al.*, 2020). Diese Form der technologischen Öffnung von Bankensystemen unter Wahrung aller Regulierungen (Datenschutz, Bankgeheimnis etc.) ermöglichen den Datenaustausch mittels Programmierschnittstellen (APIs). Dies fördert die Innovation und reduziert die Abhängigkeit von einzelnen Banken-IT-Systemen. In der Schweiz hat die SIX Group bereits 2020 die bLink Plattform gestartet um unterschiedlichsten Finanzakteuren einen einheitlichen Datenzugang zu ermöglichen.[7] Und auch der Bundesrat hat 2022 bekannt gegeben, dass er «Open Finance» fördern will und deshalb bis 2024 die Schweizer Finanzbranche die Öffnung ihrer Schnittstellen voranbringen soll (Bundesrat, 2022).

*Gesundheitswesen*

Auch im Gesundheitssektor spielen Daten und Digitaltechnologien eine wichtige Rolle. In der Schweiz sind verschiedene Initiativen im Gange, welche im Sinn der digitalen Souveränität die Technologie-Kompetenz und die digitale Selbständigkeit der Akteure stärken sollen. Bereits seit 2017 wird über das «Swiss Personalized Health Network» (SPHN) die Entwicklung von digitalen Infrastrukturen zur nationalen Nutzung von Gesundheitsdaten für Forschungszwecke vorangetrieben (Lawrence, Selter and Frey, 2020). Dies ist ein wichtiger Beitrag zur digitalen Souveränität des Schweizer Gesundheitssektors, da so neben den inhaltlichen Daten auch Rahmenbedingungen wie einheitliche Standards und Datenformate geschaffen werden (Touré *et al.*, 2023).

Bezüglich Digitalisierung des Schweizer Gesundheitswesen soll das Programm «Digisanté» im Umfang von CHF 624 Millionen von 2025 bis 2034 einen grossen Fortschritt bringen (Bundesrat, 2023b). Darin werden 54 Massnahmen und Vorhaben gebündelt, die zentrale IT-Systeme aufbauen und den Datenaustausch fördern sollen. Datensouveränität stellt dabei eines der zahlreichen Ziele des Programms dar. Als weitere Zielsetzung ist die Schaffung eines Datenraums Gesundheit vorgesehen. Unabhängig von der Vorlage wurde 2023 der Verein «Gesundheitsdatenraum Schweiz» gegründet.[8]

Auf europäischer Ebene wurde 2022 mit einem Verordnungsentwurf für einen Gesundheitsdatenraum die Initiative für einen «European Health Data Space» (EHDS) geschaffen[9]. Damit sollen die Personen mehr Kontrolle über ihre Gesundheitsdaten erhalten und gleichzeitig Innovation und Forschung im Gesundheitssektor gefördert werden. Inhaltlich wurde dies von der Wissenschaft begrüsst, jedoch auch auf das Verbesserungspotenzial bezüglich Datensouveränität durch dezentrale Datenhaltung hingewiesen (Raab *et al.*, 2023). Das Vertrauen in digital souveräne Datenhaltung und Transparenz von IT-Systemen spielen bei Gesundheitsdaten generell eine wichtig Rolle, wie Forschende über Fallstudien in England festgestellt haben (Pierri and Herlo, 2021).

---

[4] https://www.eiopa.europa.eu/digital-operational-resilience-act-dora_en

[5] https://digital-strategy.ec.europa.eu/en/policies/discover-eidas

[6] https://digital-strategy.ec.europa.eu/en/policies/eudi-wallet-implementation

[7] https://blink.six-group.com/

[8] https://gesundheitsdatenraum.ch

[9] https://health.ec.europa.eu/ehealth-digital-health-and-care/european-health-data-space_de

*Verkehrssektor*

Mobilitätsbedürfnisse und -angebote werden stetig vielseitiger, die technologische Entwicklung schafft mit E-Mobility, E-Scooters etc. laufend neue Möglichkeiten Menschen und Güter zu transportieren. So ist auch der Schienen- und Strassenverkehr in den letzten Jahren in vieler Hinsicht digitalisiert worden. Staatliche Stellen sind oftmals die Dateneigner (SBB, swisstopo etc.), wenn auch kommerzielle Angebote von US-amerikanischen Technologie-Unternehmen (bspw. Google Maps) oftmals häufiger verwendet werden als die digital souveränen Lösungen von nationalen Anbietern. Um Monopolbildungen zu vermeiden und Marktversagen vorzubeugen, empfiehlt ein Expertenbericht aus dem Jahr 2019 die Schaffung einer staatlichen Mobilitätsdateninfrastruktur (Ecoplan, 2019).

Diese Zielsetzung hatte der Bundesrat bereits vorgängig im Rahmen einer ersten Vernehmlassung zur Vorlage «Multimodale Mobilitätsdienstleistungen» aufgenommen, die Anpassungen an bestehenden Gesetzen vorgesehen hatte (UVEK, 2018). Gleichzeitig startete das Bundesamt für Verkehr verschiedene Initiativen und Projekte im Bereich der «multimodalen Mobilität», welche die integrierte Nutzung verschiedener Verkehrsmittel zum Ziel hatte (BAV, 2021). In diesem Kontext wurde u.a. der «Open Journey Planner» (OJP)[10] lanciert, ein Anbieter-neutrales Open Source Routingsystem, das ideale Mobilitätsverbindungen gemäss individueller Kriterien berechnet. Die Online-Plattform für offen zugängliche Mobilitätsdaten www.opentransportdata.swiss bietet seither eine grosse Menge an Echtzeitdaten für den Schweizer Schienen- und Strassenverkehr, die in rund 100 Anwendungen verwendet werden.[11]

Diese Beispiele zeigen auf, wie der öffentliche Sektor, Unternehmen und die Zivilgesellschaft basierend auf offenen Verkehrsdaten die Innovation und Effizienz der Mobilitätsnutzung in der Schweiz verbessern. Entscheidend dabei ist, dass sowohl die Daten als auch die Software im Besitz des öffentlichen Sektors sind und dadurch die digitale Souveränität gewährleistet ist. Durch die Freigabe als Open Government Data und Open Source Software können ausserdem Schweizer Unternehmen und zivilgesellschaftliche Organisationen die digitalen Erzeugnisse weiterentwickeln und erhöhen so auch das technologische Knowhow im Inland.

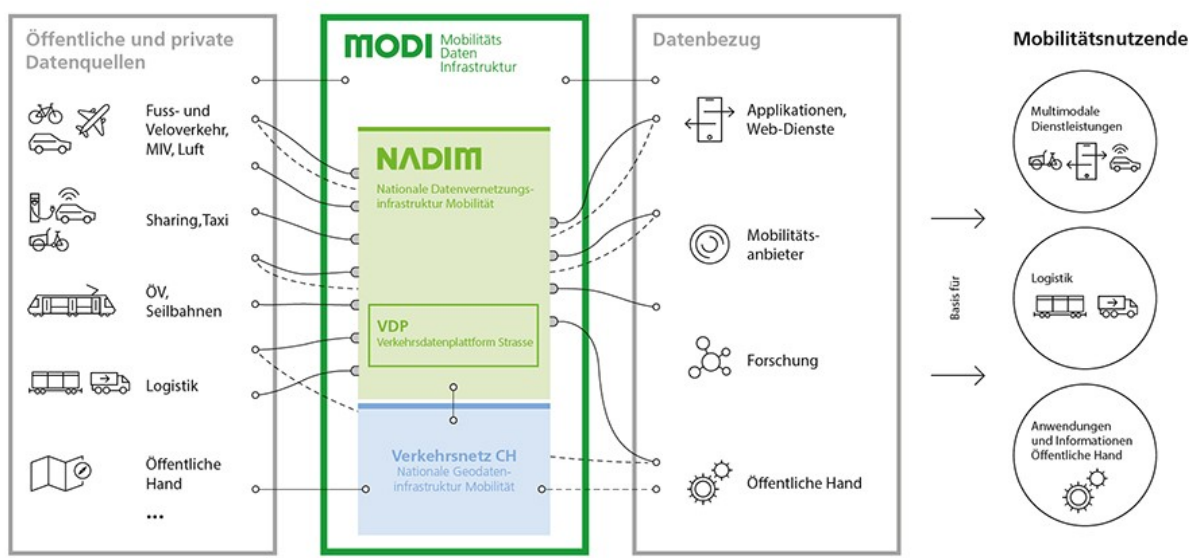

Abbildung 6: Die Mobilitätsdateninfrastruktur im Mobilitätssystem[12]

Aufgrund dieser Praxiserfahrungen und der positiven Rückmeldungen aus dem ersten Vernehmlassungsverfahren (BAV, 2020) eröffnete der Bundesrat 2022 eine zweite Vernehmlassung, dieses Mal zum neuen Gesetz für die staatliche Mobilitätsdateninfrastruktur (MODIG) und NADIM, der «Nationalen Datenvernetzungsinfrastruktur Mobilität» (UVEK, 2022). Damit will der Bundesrat die digitale Souveränität im Verkehrsdatenraum langfristig gewährleisten und Innovation im Mobilitätssektor erhöhen (siehe Abbildung 4).

---

[10] https://opentransportdata.swiss/en/cookbook/open-journey-planner-ojp/

[11] https://opentransportdata.swiss/de/showcase-5/

[12] https://www.bav.admin.ch/bav/de/home/allgemeine-themen/modi.html

Auch auf europäischer Ebene spielen Verkehrsdaten eine wichtige Rolle. So hat die Deutsche Bundesregierung bereits 2019 die Initiative für einen europäischen Mobilitätsdatenraum[13] lanciert (Pretzsch, Drees and Rittershaus, 2022). Schlüsselanliegen in diesem Kontext ist die Datensouveränität, sodass europäische Behörden, Forschungseinrichtungen und Firmen langfristig uneingeschränkten Zugang zu den Verkehrsdaten ausüben können (Drees et al., 2021).

*Bildungssektor*

Durch den vermehrten Einsatz von Digitaltechnologien im Bildungsumfeld verlagert sich auch der Unterrichtsalltag immer mehr in den digitalen Raum. So werden heute im Schulumfeld oftmals die Public Cloud Dienste von Microsoft und Google verwendet, um Dateien zu speichern, Emails zu verschicken, Online-Kurse durchzuführen etc.

Dieser Verlust an digitaler Souveränität ist deshalb der Kernpunkt einer neuen Studie aus Deutschland, welche die Problematik erläutert und Alternativen aufzeigt (Meinel, Galbas and Hagebölling, 2023). Als konkretes Beispiel wird die durch das Hasso-Plattner-Institut (HPI) initiierte «HPI Schul-Cloud» beschrieben, die seit 2017 an über 4000 Schulen von über 1.4 Millionen Nutzenden eingesetzt wird. Diese auf Open Source Software basierende IT-Umgebung, die heute unter dem Namen dBildungscloud[14] durch die Firma Dataport weiterentwickelt wird, deckt eine Vielzahl an Digitalisierungs-Bedürfnissen des Schulalltags ab (Meinel, John and Wollowski, 2022).

In der Schweiz verwenden die meisten Bildungsstellen Cloud-Lösungen von Microsoft, obwohl es datenschutzrechtlich komplex und auch umstritten ist, ob und wie die Schulen Office-Produkte und weitere Microsoft-Produkte nutzen dürfen (Kanton Basel-Stadt, 2019; Kanton St.Gallen, 2021; Kanton Zürich, 2023). Digital souveräne Alternativen bieten Open Source Lösungen wie Nextcloud, die es erlauben Daten auf Servern in der Schweiz zu speichern (Karlitschek, 2020). Nextcloud wird beispielsweise beim Open Education Server eingesetzt, der 2019 durch den Verein CH Open lanciert wurde (Stürmer, 2019) und in nächster Zeit erneuert werden soll (Stürmer and Hitz-Gamper, 2021).

Generell verhindert der Einsatz von Open Source Software im Unterricht, dass Schülerinnen und Schüler bzw. deren Eltern dazu gezwungen werden proprietäre Produkte zu kaufen – insbesondere wenn die Schulzeit vorbei ist und die Software-Tools selber beschafft werden müssen. So schreibt M. Oğuz Arslan im "European Journal of Research on Education", dass es die ethische Verantwortung von Schulen sei, Kindern digitale Fähigkeiten unabhängig von Firmenprodukten zu vermitteln (Arslan, 2014). Dennoch ist in der Schweiz Open Source Software im Bildungsumfeld wenig verbreitet, was eine aktuelle Umfrage auf gymnasialer Ebene zeigt (Taisch, 2023).

Als positives Beispiel von digitaler Souveränität kann Edulog[15] von educa.ch betrachtet werden. Dieses 2017 konzipierte E-ID Ökosystem für den Schweizer Bildungssektor (Brugger, Selzam and Buchser, 2017) wurde 2020 in Betrieb genommen und bietet einen föderierten Identitätsdienst für die Schulen an. Ebenfalls eine hohe digitale Souveränität weisen die Dienste von Switch auf, welche u.a. mit edu-ID einen Identitätsdienst für die Hochschulen bietet oder mit Switch Engines und Switch Drive selbständig Cloud-Dienste für Forschende betreibt.[16]

Der Bildungsdatenraum wurde 2019 durch educa.ch mit einem umfassenden Bericht aus konzeptioneller, rechtlicher und technologischer Sicht skizziert (educa.ch, 2019). Auf europäischer Ebene wird diese Thematik über den «Data Space Education and Skills» (DASES) im Rahmen von GAIA-X abgedeckt.[17]

---

[13] https://mobility-dataspace.eu

[14] https://dbildungscloud.de

[15] https://www.edulog.ch

[16] https://www.switch.ch/de/loesungen

[17] https://prometheus-x.org

# 5 Empfehlung konkreter Massnahmen

Wie in den vorgängigen Kapiteln erläutert, bestehen unterschiedliche Möglichkeiten für Massnahmen zur Förderung der digitalen Souveränität. In diesem letzten Abschnitt des Berichts werden Vorschläge beschrieben, wie konkret die digitale Souveränität in der Schweiz erhöht werden könnte. Die Empfehlungen richten sich weitgehenden an staatliche Stellen und zeigen Möglichkeiten auf, wie diese ihre digitale Souveränität erhöhen können.

Die beschriebenen Massnahmen haben nicht zum Ziel, eine komplett autonome Digitalisierung zu erreichen, um autarke Informatiksysteme zu betreiben. Eine solche digitale Entkoppelung widerspräche der internationalen Vernetzung der Schweiz. Vielmehr geht es bei diesen Massnahmen darum, Abhängigkeiten zu IT-Herstellern zu reduzieren und alternative Lösungen zu bieten. Dies ist auch in anderen Bereichen der Schweizer Wirtschaft der Fall, in denen keine vollständige Versorgung im Inland möglich ist.

Als Beispiel kann die Rolle der Schweiz in der Autoindustrie genannt werden: Als kleines Land hat die Schweiz keine eigenen Autohersteller und ist somit auf den Import von Fahrzeugen aus dem Ausland angewiesen. Durch die vielen ausländischen Fahrzeugproduzenten findet jedoch ein hoher Wettbewerb statt und es existiert somit eine grosse Wahlfreiheit. Ursache dafür ist der niedrige Netzwerkeffekt bei der Nutzung von Autos und deren niedrige Anbieterabhängigkeiten. So kann bei der Erneuerung der Fahrzeugflotte relativ einfach auf andere Lieferanten und Automarken gewechselt werden.

Anders ist die Situation in der IT-Industrie: Wie im Bericht aufgezeigt, führen hohe System- und Herstellerabhängigkeiten (Lock-in Effekte) in der Informatik zu eingeschränkten Wahlmöglichkeiten (siehe Seite 7 Abschnitt «System- versus Hersteller-Abhängigkeiten). Der Markt wird dadurch verzerrt, innovative Lösungen aus der Schweiz oder Europa haben grosse Eintrittshürden bzw. werden ohne staatliche Förderung gar nicht erst entwickelt. Damit stellt sich die Frage, in welcher Form der Staat eingreifen sollte, um den Wettbewerb zu stärken und die Wahlfreiheit zu erhöhen. Gezielte Massnahmen können so die Förderung der digitalen Souveränität erhöhen und damit die Behörden und die Schweizer Wirtschaft stärken.

Die nachfolgend erläuterten Massnahmen beziehen sich sowohl themenübergreifend auf die gesamte digitale Transformation in der Schweiz als auch auf die spezifischen Technologiebereiche Software, Daten, KI und ICT-Infrastruktur. Die empfohlenen Massnahmen unterstützen die Schweizer Unternehmen und Behörden bei der Schaffung von innovativen Produkten und Dienstleistungen im digitalen Sektor und erhöhen damit die digitale Souveränität in der Praxis.

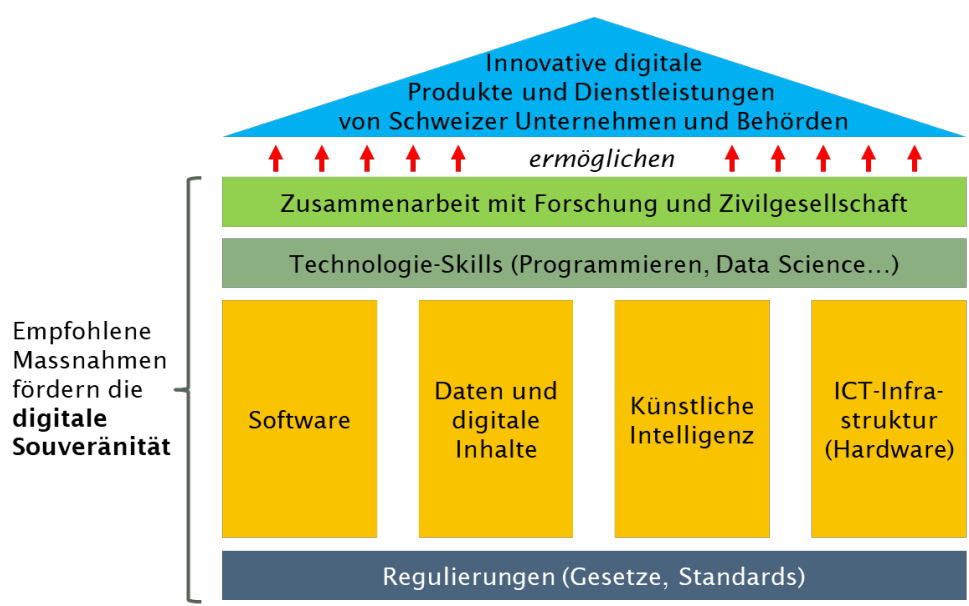

Abbildung 7: Schematische Darstellung der Aspekte der digitalen Souveränität, welche die Bereitstellung von innovativen Produkten und Dienstleistungen ermöglichen (eigene Darstellung)

Wichtig zu vermerken ist, dass in der Schweiz bereits zahlreiche Aktivitäten auf politischer, verwaltungsbezogener und wissenschaftlicher Ebene stattfinden. Es gibt verschiedene Organisationen und politische Forderungen, die bereits an der Stärkung der digitalen Souveränität wirken. Nun braucht es eine Bündelung der vorhandenen Kräfte, eine Vernetzung der engagierten Akteure, gezielte Finanzierung von Massnahmen und eine zentrale Koordination der Umsetzung.

Bezüglich Kosten ist pro Massnahme eine grobe Schätzung angegeben, welche finanziellen Mittel einmalig und/oder wiederkehrend zur erfolgreichen Umsetzung investiert werden müssten. Diese Angaben sind verständlicherweise mit einer sehr hohen Ungenauigkeit behaftet bzw. können die Massnahmen in der Regel auch mit wesentlich höheren oder geringeren Beiträgen realisiert werden.

Die angegebenen Zuständigkeiten zeigen mögliche Stellen in der Bundesverwaltung auf, welche die jeweilige Massnahmen verantworten und realisieren:

- BAV: Bundesamt für Verkehr
- BFS: Bundesamt für Statistik
- BIT: Bundesamt für Informatik und Telekommunikation
- BJ: Bundesamt für Justiz
- DTI: Bundeskanzlei, Bereich Digitale Transformation und IKT-Steuerung
- DVS: Digitale Verwaltung Schweiz
- SBFI: Staatssekretariat für Bildung, Forschung und Innovation

### 5.1 Themenfeld-übergreifende Empfehlungen

Die folgenden Empfehlungen betreffen Massnahmen, die notwendig sind, um die digitale Souveränität in der Schweiz unabhängig von den jeweiligen technologischen Themenfeldern zu fördern:

*Massnahme 1: eCH-Standard für «digitale Souveränität» schaffen*

*Schätzung einmalige Kosten: CHF 500'000*
*Mögliche Zuständigkeit: DTI und Verein eCH*

Die klare Definition des Begriffs «digitale Souveränität» ist wichtig, um ein gemeinsames Begriffsverständnis zu schaffen und deren praktische Anwendung zu ermöglichen. So bedarf es einer Operationalisierung von «digitaler Souveränität», um diese beispielsweise als Kriterium bei IT-Beschaffungen zu verwenden, um ähnlich lautende Angebote zu vergleichen, um Marktentwicklungen zu analysieren, um Audits von digitalen Lösungen durchzuführen oder um Zertifizierungen von bestimmten Produkten und Dienstleistungen zu ermöglichen.

Die Schaffung eines eCH-Standards für digitale Souveränität ist möglich und sinnvoll, da die Organisation seit über 20 Jahren sowohl durch den öffentlichen Sektor als auch von der Privatwirtschaft getragen wird und breit akzeptiert ist (Alabor and Frey-Eigenmann, 2021). Auch wurde an einer eCH-Abendveranstaltung im September 2023 die Thematik bereits eingehend diskutiert (Rutz, 2023).

Inhaltlich sollten einerseits generelle Vorgaben für digitale Souveränität von digitalen Gütern festgelegt werden, die beispielsweise die uneingeschränkte Verfügbarkeit der Technologie auf dem Internet, eine offene Lizenz, Qualitätsstandards, fundierte Dokumentation und eine aktive Fachgemeinschaft (Community) voraussetzt. Andererseits sollten je nach Art der digitalen Güter spezifische Kriterien vorgegeben werden, so muss zum Beispiel bei Software eine hohe Benutzerfreundlichkeit, Performance, Sicherheit, API-Schnittstellen etc. vorhanden sein. Oder bei Daten müssen Metadaten vorhanden sein, Dateninteroperabilität muss gewährleistet werden etc. Und bei KI-Modellen sollten die Trainingsdaten bekannt und die Modellparameter und Hyperparameter zugänglich sein etc.

*Massnahme 2: Knowhow-Aufbau und Weiterbildungsmöglichkeiten*

*Schätzung wiederkehrende Kosten: CHF 1 bis 10 Millionen pro Jahr*
*Mögliche Zuständigkeit: SBFI*

Wie in den obenstehenden Beispielen aus der Schweiz und im Ausland aufgezeigt, bedarf es fundierten technischen Skills, um IT-Systeme selbständig zu betreiben, gute Software zu entwickeln, zuverlässige Datenanalysen vorzunehmen und komplexe KI-Modelle zu trainieren. Der Betrieb von Servern erfordert Erfahrung in der Systemadministration und IT-Security, Software-Entwicklung benötigt Kenntnisse in den jeweiligen Programmiersprachen, Data Science Wissen ist erforderlich für Datenanalysen und mathematische Grundlagen sowie praktische Informatik-Fähigkeiten sind notwendig für die Erstellung von KI-Modellen.

Obschon dies anspruchsvolle Aufgaben sind, werden sie doch täglich tausendfach in der Schweiz praktiziert. Somit wird kein seltenes Geheimwissen benötigt, sondern es braucht einfach möglichst viele kompetente Mitarbeitende im Informatiksektor, die ihr Wissen laufend aktualisieren. Aus diesem Grund ist es wichtig, dass sowohl das verwaltungsinterne IT-Knowhow und als auch die «Digital Skills» in der breiten Bevölkerung erhöht werden. Diese Zielsetzung wird bspw. durch die Schaffung von Anreizen für entsprechende Aus- und Weiterbildungsangebote sowie durch die Bereitstellung von fundierten Online-Lernangebote umgesetzt. Auch die stärkere Förderung von MINT-Berufen (Mathematik, Informatik, Naturwissenschaften und Technik) auf allen Bildungsstufen leistet einen wesentlichen Beitrag zur digitalen Souveränität, in dem mehr junge Menschen in der Schweiz die notwendigen technischen Fähigkeiten erlernen. Des Weiteren ist die Entlöhnung ein wichtiger Faktor. Mit Salären der grossen IT-Konzerne können Behörden nicht mithalten. Dennoch gibt es viele kompetente Fachleute, die bereit sind auf ein sehr hohes Einkommen zu verzichten, um dafür einen Dienst für die Öffentlichkeit zu erbringen. Wichtig ist dabei ein innovatives, fehlertolerantes Behördenumfeld, das ähnlich attraktiv ist wie bspw. die Arbeitsumgebung der «Hyperscalers».

*Massnahme 3: Zusammenarbeit der Verwaltung mit Forschung und Zivilgesellschaft*

*Schätzung wiederkehrende Kosten: CHF 1 Million pro Jahr*
*Mögliche Zuständigkeit: DTI und BIT*

Der technologische Fortschritt geschieht rasant, deshalb ist der Staat darauf angewiesen, dass er sich regelmässig mit der IT-Fachwelt austauscht. Traditionellerweise geschieht dies über Partnerschaften oder öffentliche Beschaffungen mit der Privatwirtschaft. Dies macht auch Sinn, denn die IT-Industrie entwickelt und betreibt Digitaltechnologien als deren Hauptaufgabe. Dennoch wollen und müssen Firmen letztlich auf Umsatz und Profit achten, weshalb Zielkonflikte mit dem staatlichen Handeln offensichtlich sind. Der Privatsektor hat den «Shareholder Value» zum Ziel, während der öffentliche Sektor den «Public Value» als höchstes Prinzip verfolgt.

Eine Alternative bietet die Zusammenarbeit mit der Wissenschaft und zivilgesellschaftlichen Organisationen. So besitzt die Schweiz hervorragende Hochschulen, die national und international erfolgreich sind. Forschende in der Informatik, Wirtschaftsinformatik, Computerlinguistik, Statistik und anderen MINT-Fächern besitzen wertvolles Digitalisierungs-Knowhow, das sie bereits heute unter anderem für staatliche Vorhaben anwenden. Die Förderung der Zusammenarbeit zwischen Behörden und Digitalisierungsforschenden stärkt die digitale Souveränität durch vermehrten Knowhow-Transfer von der Wissenschaft in die Praxis und umgekehrt.

Eine weitere wichtige Ressource sind Vereine und Communities in der Schweiz, die sich um spezifische Digitaltechnologien formiert haben. So existieren dedizierte, nicht gewinnorientierte Organisationen bspw. zur freien Geodatenplattform OpenStreetMap (OpenStreetMap Schweiz[18]), zur Open Source Datenbank PostgreSQL (Swiss PostgreSQL Users Group[19]) oder zum Open Source Content Management System Drupal (Drupal Switzerland[20]). Diese Vereine organisieren regelmässige interne Meetings zum Wissens- und Erfahrungsaustausch und sind oft auch an der Durchführung von nationalen Anlässen beteiligt. Durch die Beteiligung von Mitarbeitenden aus dem öffentlichen Sektor oder durch die Kooperation mit den entsprechenden Communities (siehe bspw. Hitz-Gamper and Stürmer, 2021) geschieht ein wertvoller Wissenstransfer, der schliesslich die digitale Souveränität von Schweizer Behörden stärkt.

---

[18] https://osm.ch

[19] https://www.swisspug.org

[20] https://drupal.ch

**5.2 Software**

Diese Empfehlungen betreffen Massnahmen, welche die digitale Souveränität bei der Software erhöht:

*Massnahme 4: Förderung Open Source Software und Open Standards im öffentlichen Sektor*

*Schätzung wiederkehrende Kosten: CHF 1 Million pro Jahr*
*Mögliche Zuständigkeit: DTI oder BIT*

Wie in den vorangehenden Kapiteln detailliert erläutert, reduziert der Einsatz und die Freigabe von Open Source Software die Hersteller-Abhängigkeiten (siehe Seite 7, Abschnitt 1.3 «System- versus Hersteller-Abhängigkeiten») und erhöht damit die digitale Souveränität signifikant. Des Weiteren stärkt auch die Durchsetzung von offenen Datenformaten und die Verfügbarkeit von APIs die Interoperabilität und damit die digitale Souveränität (siehe Seite 21, Abschnitt «Software»). Deshalb wird empfohlen, Open Source Software, Open Standards und offene Schnittstellen im öffentlichen Sektor mit neuen Leitfäden und personellen Mitteln zu fördern.

Im Jahr 2019 wurden durch das damalige Informatiksteuerungsorgan des Bundes (ISB) ein «Strategischer Leitfaden Open Source Software in der Bundesverwaltung» (ISB, 2019b) und ein «Praxis-Leitfaden Open Source Software in der Bundesverwaltung» (ISB, 2019a) publiziert. Diese Dokumente sehen vor, dass weitere Massnahmen ergriffen werden, um den Einsatz und die Veröffentlichung von Open Source Software in der Bundesverwaltung zu fördern.

Ein wichtiges Element sollte ein neuer Leitfaden zur fachgerechten Freigabe von Open Source Software durch Behörden darstellen. Gemäss EMBAG (siehe Seite 12, «Möglichkeiten des EMBAG zur Reduktion von Herstellerabhängigkeiten») müssen ab 2024 grundsätzlich alle Software-Entwicklungen des Bundes unter einer Open Source Lizenz freigegeben werden. Dies bedingt entsprechendes Wissen, wie bei solchen Open Source Publikationen vorgegangen werden muss.

Wichtig bei IT-Beschaffungen sind ausserdem Empfehlungen und Good Practices, wie die Vorgaben des EMBAG bereits bei der Ausschreibung korrekt umgesetzt werden. So benötigt die Beschaffung von Open Source Software vertieftes Knowhow, weshalb diesbezüglich ein weiterer Leitfaden rund um die Beschaffung von Software-Applikationen und Individual-Entwicklungen unterstützen sollte.

Langfristig bringt der Aufbau eines Open Source Programme Office (OSPO) Stabilität und Professionalität in den Umgang mit Open Source Software in der Schweizer Bundesverwaltung, so wie dies die EU und Frankreich bereits realisiert haben (siehe Seite 14 «Frankreich» und Seite 15 «Open Source Strategie der EU»). Dieses OSPO kann mit zwei Vollzeitstellen die Aktivitäten der Bundesverwaltung rund um Open Source Software koordinieren, Fragen beantworten, eine interne «Community of Practice» leiten und Vorhaben vorantreiben (McAffer, 2019; OpenForum Europe, 2022). Und ähnlich wie das «Zentrum digitale Souveränität» in Deutschland könnte das Bundes-OSPO in der Schweiz auch übergreifende Anliegen rund um digitale Souveränität wahrnehmen. Eine enge Zusammenarbeit mit der Digitalen Verwaltung Schweiz (DVS) ist empfehlenswert, denn auch die Kantone und Gemeinden veröffentlichen regelmässig Open Source Software und haben diesbezüglich Fragen aus der Praxis.

*Massnahme 5: Plattform für die Freigabe von Behörden-Anwendungen*

*Schätzung einmalige Kosten: CHF 1 Million, Schätzung wiederkehrende Kosten: CHF 200'000 pro Jahr*
*Mögliche Zuständigkeit: DTI oder BIT, Zusammenarbeit mit DVS*

Durch die Schaffung einer nationalen Plattform für die Freigabe von Behörden-Anwendungen kann ein zentraler Einstiegspunkt («Single Point of Entry») für Open Source Software von und für den öffentlichen Sektor geschaffen werden. Wie oben beschrieben, wurde so eine Plattform bereits in Deutschland und Frankreich realisiert und auch die EU betreibt mit OSOR einen entsprechenden Dienst. So könnte in der Schweiz, ähnlich wie opendata.swiss für Open Government Data von Bund, Kantonen und weiteren staatlichen Stellen, auch für Open Source Software ein nationales Quellcode-Portal aufgebaut und bekannt gemacht werden.

Damit würden Gemeinden und Kantone beim Finden und Integrieren von bereits vorhandenen Open Source Lösungen unterstützt. Ausserdem könnten Anbieter für Dienstleistungen rund um die entsprechende Open Source Software ihre Fähigkeiten und Success Stories präsentieren. Dies würde die Marktentwicklung fördern, die Abhängigkeit von IT-Firmen reduzieren und so die digitale Souveränität stärken.

**5.3 Daten**

Die folgenden Empfehlungen betreffen Massnahmen, welche die digitale Souveränität bezüglich Datenspeicherung und Datenverarbeitung erhöhen:

*Massnahme 6: Mobilitätsdateninfrastrukturgesetz (MODIG) vorantreiben*

*Schätzung der Kosten: keine zusätzlichen Ausgaben (Gesetzgebungsprozess)*
*Zuständigkeit: BAV*

Das Mobilitätsdateninfrastrukturgesetz (MODIG) will den Datenaustausch zwischen Infrastrukturbetreibern, Verkehrsunternehmen, privaten Anbietern und Verkehrsteilnehmenden fördern und dazu ein entsprechendes Gesetz und eine zentrale, staatliche IT-Plattform einführen (siehe auch Seite 25, Abschnitt «Verkehrssektor»). Wie im erläuternden Bericht der Vernehmlassung ausgeführt, soll so die rechtliche Grundlage geschaffen werden, dass der öffentlichen und private Verkehr die physische Infrastruktur in der Schweiz besser ausnützen kann (UVEK, 2022).

Auch soll mit dem MODIG die Innovation gefördert und neue Geschäftsfelder für Mobilitätsanbieter geschaffen werden. Damit kann einem Marktversagen vorgebeugt werden, da kein Akteur allein die notwendigen Ressourcen für den Aufbaue der IT-Infrastruktur wird. Ausserdem kann durch die staatliche Umsetzung die digitale Souveränität gewährleistet werden, da so keine Abhängigkeiten zu ausländischen Technologie-Firmen aufgebaut wird. Es ist somit empfohlen, diesen Gesetzgebungsprozess voranzubringen, um das Potenzial der Datennutzung im Mobilitätssektor zu realisieren.

*Massnahme 7: Sekundärnutzung von Daten auf Interoperabilität ausrichten*

*Schätzung der Kosten: keine zusätzlichen Ausgaben (Gesetzgebungsprozess)*
*Zuständigkeit: BJ*

Wie im Abschnitt über die aktuellen Regulierungen der EU erwähnt, wird zur Zeit der EU Data Act verhandelt, der die Dateninteroperabilität erhöht und so mehr Wettbewerb in die Datenökonomie bringen soll. Die Meinungen diesbezüglich gehen auseinander, aber eine Mehrheit der Unternehmen aus der Industrie scheint die neue Regulierung zu begrüssen (Böhler, 2024).

Auch in der Schweiz soll durch die Mehrfachnutzung von Daten das ökonomische Potenzial der Digitalisierung besser genutzt werden. So will der Bundesrat bis Ende 2024 eine zentrale Anlaufstelle «Datenökosystem Schweiz» für jährlich CHF 1.4 Millionen aufbauen (Bundesrat, 2023a). Das Vernehmlassungsverfahren für ein neues Rahmengesetz für die Sekundärnutzung von Daten soll per Ende 2026 gestartet werden. Damit soll Rechtssicherheit bezüglich Datenschutz, Urheberrechten und weiteren juristischen Aspekten geschaffen werden.

Dies sind positive Entwicklungen, welche die Datensouveränität und damit auch die digitale Souveränität fördern. Dennoch besteht die Gefahr, dass nicht das gesamte Potenzial genutzt wird. Währenddem in der EU die Dateninteroperabilität und die Datenfreigabe auch von privaten Akteuren eingefordert wird, soll dies in der Schweizer Gesetzgebung aufgrund der Wirtschaftsfreiheit nicht unbedingt der Fall sein. Deshalb ist empfohlen, dass auch in der Schweiz Unternehmen und andere Datenbesitzer dazu verpflichtet werden können ihre Daten freizugeben, falls dies der Gesellschaft und der Volkswirtschaft nützt. Wichtig ist ausserdem die Freigabe von Daten in offenen Formaten und Standards, sodass für die Datenverarbeitung keine proprietären Applikationen notwendig sind und bei einem Systemwechsel die Daten tatsächlich weiterverwendet werden können.

*Massnahme 8: Plattform für Speicherung und Freigabe von Open Government Data (OGD)*

*Schätzung einmalige Kosten: CHF 1 Million*
*Schätzung wiederkehrende Kosten: CHF 200'000 pro Jahr*
*Mögliche Zuständigkeit: BFS, in Zusammenarbeit mit DVS*

Heute schon werden viele Behördendaten als Open Government Data (OGD) von zahlreichen staatlichen Stellen auf opendata.swiss veröffentlicht. Mit dem EMBAG Artikel 10 wird dies noch zunehmen, da ab 2024 (mit einer Übergangsfrist von drei Jahren) alle Bundesstellen grundsätzlich alle ihre Daten freigeben müssen (Bundesversammlung, 2023). Allerdings ist opendata.swiss 'lediglich' ein Metadatenportal. Das heisst, dass dort bloss Beschreibungen und Links veröffentlicht sind, die eigentlichen Daten werden weiterhin auf den Servern der Behörden gespeichert. Dies führt dazu, dass es bspw. für kleinere Gemeinden sehr aufwändig ist, eigene Datenbestände freizugeben.

Deshalb ist empfohlen, dass der Bund eine zentrale Online-Plattform aufbaut, die sowohl Daten speichert und über jene auch Daten freigegeben werden können. Diese muss gut integriert sein mit bestehenden Dateninitiativen rund um OGD, der nationalen Datenbewirtschaftung (NaDB), der Interoperabilitätsplattform (I14Y) und dem Linked Data Service LINDAS (Wüst, 2022).

### 5.4 IT-Infrastruktur

Diese Empfehlungen umfassen Massnahmen, welche die digitale Souveränität auf der Ebene der IT-Infrastruktur stärken:

*Massnahme 9: Aufbau der Swiss Government Cloud basierend auf Open Source Technologien*

*Schätzung der Kosten: rund CHF 300 Millionen*
*Zuständigkeit: BIT*

Wie bereits erwähnt, ist das BIT an den Vorbereitungen des Aufbaus einer «Swiss Government Cloud» für die öffentlichen Verwaltungen der Schweiz. Diese Initiative ist zu unterstützen, denn so wird die digitale Souveränität bezüglich der IT-Infrastruktur und der Datenkontrolle deutlich verbessert (siehe Seite 9 «Trade-off zwischen Public Cloud Services und digitaler Souveränität»). Gleichzeitig muss nicht alles neu entwickelt werden, sondern kann auf Erfahrungen mit dem «Sovereign Cloud Stack» aus Deutschland zurückgreifen (siehe Seite 13 «Sovereign Cloud Stack»). Entsprechend ist es wichtig, dass das BIT sich mit den bereits vorhandenen Open Source Cloud Technologien befasst.

*Massnahme 10: «Software-as-a-Service» (SaaS) Angebote für Schweizer Behörden anbieten*

*Schätzung einmalige Kosten: CHF 2 Millionen*
*Schätzung wiederkehrende Kosten: CHF 500'000 pro Jahr*
*Mögliche Zuständigkeit: BIT*

Das in Massnahme 9 beschriebene Szenario erhöht die digitale Souveränität auf der technologisch tiefsten Stufe, der «Infrastructure-as-a-Service» (IaaS). Höhere Ebenen wie «Platform-as-a-Service» (PaaS) und «Software-as-a-Service» (SaaS) sind jedoch ebenfalls wichtig zur Erhöhung der digitalen Souveränität, da sich auf diesen Stufen typischerweise die Geschäftslogik befindet und die Datenverarbeitung durchgeführt wird. Ausserdem sind insbesondere SaaS-Lösungen für die Endbenutzer sichtbar, was die Attraktivität und Sichtbarkeit von solchen «Private Cloud Services» erhöht (Hentschel and Leyh, 2018).

Heute gibt es viele praktische Cloud-Angebote wie Doodle, Slack, Microsoft Teams, Google Drive oder Dropbox, die auch von Schweizer Verwaltungsstellen im Alltag oft benutzt werden. Allerdings steht hinter allen Cloud-Services ein Geschäftsmodell, das entweder auf Werbeeinnahmen basiert oder bei dem durch Herstellerabhängigkeiten hohe Lizenzpreise bezahlt werden müssen. Auch ist es unklar, ob diese vorwiegend von amerikanischen IT-Unternehmen betriebenen SaaS-Lösungen mit der Europäischen bzw. der Schweizerischen Datenschutzgesetzgebung kompatibel sind. Und zuletzt ist es auch aus Sicht der digitalen Souveränität kritisch, wenn Anwendungen und Daten des öffentlichen Sektors in der Kontrolle von kommerziellen, ausländischen IT-Anbietern liegen.

Deshalb wird empfohlen, entsprechende Open Source Alternativen zu proprietären Lösungen zu identifizieren, bspw. Nuudel für Doodle, Elements für Slack, BigBlueButton für Microsoft Teams sowie Nextcloud für Google Drive und Dropbox. Diese Open Source Anwendungen können auf Servern des BIT installiert und für den öffentlichen Sektor zur Verfügung gestellt werden. Solche Applikationen ermöglichen insbesondere kleineren Behörden im Alltag den niederschwelligen Zugang zu Cloud-Services, bei denen die Daten digital souverän auf der IT-Infrastruktur der Bundesverwaltung bleiben.

Ausserdem bietet es sich an, den durch das ZenDiS entwickelte openDesk (siehe Seite 13 Abschnitt «openDesk als digital souveräner IT-Arbeitsplatz») als digital souveränen Online-Arbeitsplatz von Bundesstellen, kantonalen Behörden und kommunalen Verwaltungen anzubieten.

*Massnahme 11: Cloud-Lösung für internationale Organisationen anbieten*

*Schätzung der Kosten: keine Angaben möglich*
*Mögliche Zuständigkeit: BIT*

Die Schweiz kann durch ihre politische Neutralität und hohe Digitalkompetenz der Bundesinformatik eine einzigartige Rolle als digitaler Gaststaat wahrnehmen und eine rechtlich und technisch digital souveräne Cloud für internationale Organisationen anbieten. Basierend auf den Erfahrungen und den

Technologien mit der Swiss Government Cloud (Massnahme 9) und den SaaS-Angeboten für Schweizer Behörden (Massnahme 10) kann das BIT eine weitere Cloud-Umgebung für globale, multilaterale Institutionen aufbauen. Wie im Abschnitt zum IKRK beschrieben (siehe Seite 18, Abschnitt 3.8), haben diese Organisationen spezifische Anforderungen für eine sichere und vertrauenswürdige IT-Infrastruktur.

Wichtig ist für diese internationalen Organisationen die «digitale Immunität», sodass Daten und IT-Systeme vollständig eigenständig genutzt werden können, ohne dass technische oder juristische Abhängigkeiten zu IT-Firmen und deren Ursprungsländer entstehen. Auf der Basis von Open Source Software können solche umfassend digital souveränen Cloud-Lösungen aufgebaut und betrieben werden. Dabei sollen Daten einerseits sicher verarbeitet und abgelegt werden können. Andererseits sollen auch geschäftskritische Applikationen betrieben und mittels eigenständiger Modelle (siehe Massnahme 12) die neusten Errungenschaften der künstlichen Intelligenz genutzt werden.

### 5.5 Künstliche Intelligenz

Die folgenden Empfehlungen betreffen Massnahmen, um die digitale Souveränität im Bereich der künstlichen Intelligenz zu stärken:

*Massnahme 12: Anpassungen und Betrieb eigener KI-Modelle*

*Schätzung einmalige Kosten: CHF 2 Millionen*
*Schätzung wiederkehrende Kosten: CHF 500'000 pro Jahr*
*Mögliche Zuständigkeit: BIT oder BFS*

Angebote der künstlichen Intelligenz wie ChatGPT können heute über das Web-Interface (Browser) oder über ein «Application Programming Interface» (API) bequem angewendet werden. Allerdings wird durch die regelmässige Nutzung eine hohe Abhängigkeit zum IT-Anbieter aufgebaut, da dieser die KI-Lösung vollständig kontrolliert und sämtliche verwendeten Technologien und Trainingsdaten intransparent bleiben. Auch werden die Befehle («Prompts») und die Nutzungsdaten von den Anwendenden an die Anbieter von KI-Diensten übermittelt wobei unklar bleibt, was damit geschieht. Die Alternative ist die Entwicklung und der Betrieb von eigenen Sprachmodellen basierend auf Open Source KI Modellen (siehe dazu Abschnitt «Künstliche Intelligenz»).

Das initiale Erstellen («Pre-Training») von grossen Sprachmodellen («Large Language Models» oder «Foundation Models») benötigt sehr grosse Rechenkapazitäten und riesige Datenmengen an Text, sodass heute oftmals nur grosse IT-Konzerne oder sehr spezialisierte KI-Unternehmen solche Modelle von Grund auf entwickeln können (Bender *et al.*, 2021). Verfahren zur Anpassung von KI-Modellen («Fine-Tuning») ermöglichen jedoch eine ressourcenschonende Integration bestehender Sprachmodelle für die benötigte Anwendung (Devlin *et al.*, 2018). So können auch Schweizer Behörden Modelle für ihre Bedürfnisse anpassen und produktiv einsetzen. Beispielsweise verwendet das Bundesgericht eines auf die Schweizer Landessprachen angepasstes Rechtssprachemodell für die Anonymisierung seiner jährlich rund 7000 Gerichtsurteilen (Niklaus *et al.*, 2023). Heute können Hunderttausende von KI-Modellen auf der Plattform Hugging Face[21] heruntergeladen, bei Bedarf angepasst und anschliessend auf eigenen Servern produktiv genutzt werden.

Es wird deshalb empfohlen, dass der Bund das technische Knowhow und die notwendige IT-Infrastruktur aufbaut um selbst KI-Modelle trainieren und betreiben zu können. So kann Wissen und Erfahrung gesammelt werden, wie Sprachmodelle selber entwickelt und angepasst werden, was die digitale Souveränität in der künstlichen Intelligenz gegenüber dem blossen Gebrauch von ChatGPT deutlich erhöht. Insbesondere durch die enge Zusammenarbeit mit Schweizer Hochschulen kann der öffentliche Sektor von dem vorhandenen Fachwissen und der IT-Infrastruktur ideal profitieren.

Konkret könnte die Bundesverwaltung als Übersetzungslösung anstelle bzw. zusätzlich zu Deepl (Bundeskanzlei, 2019; Christian Wingeier, 2023) auch eigens angepasste Sprachmodelle verwenden. So würden einerseits Datenschutzprobleme gelöst, die durch die externe Übermittlung von vertraulichen Texten entstehen. Auf eigens betriebenen KI-Modellen würden die Daten stets auf Servern der Bundesverwaltung bleiben. Andererseits könnte der Bund bei eigenen Sprachmodellen auch ideal seine umfassende Terminologie-Datenbank (siehe TERMDAT[22]) integrieren und so die konsistente Begriffsverwendung automatisiert sicherstellen.

---

[21] https://huggingface.co/models

*Massnahme 13: Nutzung von nationaler KI-Infrastruktur für Open Source KI-Modelle*

*Schätzung der Kosten: keine Angaben möglich*
*Zuständigkeit: ETH Zürich und ETH Lausanne*

Neben IT-Konzernen und KI-Firmen haben in den letzten Jahren auch staatliche Forschungsstellen sogenannte «High Performance Computing» Infrastrukturen erneuert, um diese für die Generierung von «Foundation Models» zu nutzen. So betreibt beispielsweise Finnland mit dem LUMI Supercomputer[23] den grössten europäischen Forschungscomputer. Damit haben Forschende im Dezember 2023 anspruchsvolle Datenexperimente durchführen können (Muennighoff *et al.*, 2023).

Auch in der Schweiz soll 2024 ein neuer Hochleistungsrechner im Swiss National Supercomputing Centre (CSCS) in Lugano in Betrieb genommen werden, der möglicherweise noch leistungsfähiger als der finnische LUMI Supercomputer sein wird. Dieser Schweizer Hochleistungsrechner namens «Alps» wird 10'000 der neusten Chip-Generation von NVIDIA enthalten, den Grace-Hopper-Superchip, der CPUs und GPUs vereinigt (Fulterer and Titz, 2023). Damit werden Schweizer Forschende unabhängig von privaten Technologie-Konzernen neue «Foundation Models» erstellen können, was die digitale Souveränität stärkt.

In der vorliegenden Ankündigung weisen die ETH Zürich und die ETH Lausanne explizit darauf hin, dass der neue Schweizer Supercomputer insbesondere frei verfügbare Open Source KI-Modelle generieren und so zur digitalen Souveränität beitragen soll. In diesem Zusammenhang soll die neue «Swiss AI Initiative» den Forschenden in der Schweiz auch den Zugang zu dieser neuen KI-Infrastruktur ermöglichen (ETH Zurich and EPFL, 2023). Es wird deshalb empfohlen, die geplante Ausrichtung des neuen Schweizer Supercomputers zur Förderung der digitalen Souveränität mittels Open Source KI Modellen konsequent umzusetzen.

---

[22] https://www.bk.admin.ch/bk/de/home/dokumentation/sprachen/termdat.html

[23] https://www.lumi-supercomputer.eu

# Literaturverzeichnis